\documentclass[10pt,aps,prl,twocolumn,superscriptaddress,floatfix]{revtex4-2}

\usepackage[utf8]{inputenc}
\usepackage[colorlinks=true,allcolors=blue]{hyperref}
\usepackage{amsfonts}
\usepackage{dsfont}
\usepackage{mathtools}
\usepackage[capitalise]{cleveref}
\usepackage{graphicx}
\usepackage{physics}
\usepackage{xcolor}
\usepackage{cancel}
\usepackage{bm}
\usepackage{MnSymbol}

\usepackage[caption=false]{subfig}

\newcommand{\tzinner}[2]{\langle #1 \mid \Sigma \mid #2 \rangle}

\newcommand{\An}{\mathcal{A}_n}
\newcommand{\gnn}{g_{n}}
\newcommand{\tw}{\tilde{w}}
\newcommand{\unk}{\psi_{n}(k)}
\newcommand{\dkunk}{\partial_k \psi_{n}(k)}

\newcommand{\xc}{x_c}
\newcommand{\kc}{k_c}
\newcommand{\En}{\omega_n}

\newcommand{\bbraket}[1]{\llangle #1 \rrangle}

\begin{document}

\title{Magnon dynamics driven by the symplectic quantum metric in a chiral soliton lattice}

\author{Sidhartha Chatterjee}
\email{sidhartha.chatterjee@uni.lu}

\author{Andreas Haller}
\affiliation{Department of Physics and Materials Science, University of Luxembourg, 1511 Luxembourg, Luxembourg}

\author{Peter P. Orth}
\affiliation{Department of Physics and Center for Quantum Technologies (QuTe), Saarland University, 66123 Saarbrücken, Germany}

\author{Thomas L. Schmidt}
\email{thomas.schmidt@uni.lu}
\affiliation{Department of Physics and Materials Science, University of Luxembourg, 1511 Luxembourg, Luxembourg}

\date{\today}

\begin{abstract}

We investigate the effects of quantum geometry on the motion of magnon wave packets in one-dimensional quantum spin chains.  
In contrast to fermionic systems, where transport phenomena driven by the quantum metric have been extensively studied, we show that the dynamics in bosonic magnon systems are strongly affected by the symplectic structure and the pseudo-unitarity of the transformation of the magnon Bogoliubov–de Gennes Hamiltonian. A chiral soliton lattice provides a setting in which the symplectic quantum metric strongly influences the longitudinal magnon dynamics. We derive the semiclassical response of a magnon wave packet in such a system to a weak, slowly varying perturbation and demonstrate that the wave packet velocity acquires a geometric contribution proportional to the momentum derivative of the symplectic quantum metric, which dominates over the conventional group velocity for a nearly flat band. Solving the resulting equations of motion, we obtain a bounded trajectory in real time with a phase-space area determined by the symplectic quantum metric.
\end{abstract}

\maketitle

\textit{Introduction.} Quantum geometry provides a unified description of how quantum states change locally over parameter space. This information is conveniently encoded in the quantum geometric tensor (QGT), whose antisymmetric part is the Berry curvature, which characterizes the local twist of the Bloch states and plays a central role in the topology and dynamics of electronic as well as magnetic insulators \cite{PhysRevLett.88.207208, PhysRevLett.106.197202, PhysRevB.89.054420, doi:10.1126/science.abe4441}. The symmetric component of the QGT, the quantum metric \cite{Provost1980}, provides a measure of the distance between Bloch states in parameter space. It has gained relevance in condensed-matter physics as a tool to understand correlated phenomena such as flat-band superconductivity \cite{Peotta2015, PhysRevB.96.064511}, fractional Chern insulators \cite{PhysRevResearch.6.L032063}, moiré systems \cite{PhysRevResearch.5.L012015, Törmä2022}, nonlinear transport \cite{Gao2023, doi:10.1126/science.adq3255}, and localization \cite{nbn5-lznf}. Recent measurements of the quantum metric in electronic systems \cite{doi:10.1126/science.ado6049, Jin2026, Kang2025} have opened new avenues for direct experimental access to the quantum geometry.

Most dynamical manifestations of quantum geometry have been studied in two- and three-dimensional systems, where the Berry curvature generates a transverse anomalous velocity. In a strictly one-dimensional system, the absence of transverse motion means that an intrinsic momentum-space Berry curvature does not exist. In contrast, the quantum metric can remain nonzero even in one dimension and exhibit a strong momentum dependence. One-dimensional systems, therefore, provide a platform where longitudinal responses generated by the quantum metric can be separated from conventional Berry-curvature effects. Although the quantum geometry of bosonic Bogoliubov bands has recently attracted attention \cite{1sz1-nwzf}, its consequences for longitudinal wave-packet dynamics in one-dimensional textured magnets remain largely unexplored.

In this Letter, we demonstrate that a one-dimensional chiral soliton lattice, which exists as the classical ground state of a magnet with competing Heisenberg exchange and Dzyaloshinskii-Moriya interactions (DMI), can serve as a platform for magnon quantum-geometry effects. The nonlinear spin texture effectively transforms a uniform external field into a periodic, position-dependent potential for magnonic excitations. This results in a series of gapped bound states described by the Pöschl-Teller potential \cite{KISHINE20151}. The key finding of our work is the wave-packet dynamics driven by the symplectic quantum metric and its consequences in nearly flat magnon bands. Due to the momentum dependence of the wavefunction, the metric correction dominates over the group velocity. By preserving bosonic commutation relations through a pseudo-unitary transformation, we derive the wave-packet dynamics in the presence of a position-dependent perturbing potential with symplectic quantum-metric correction. Our results establish quasi-1D chiral soliton lattices as a promising platform to investigate quantum-metric-driven transport of bosonic excitations.

Recent studies using Lorentz microscopy have revealed the stabilization of chiral soliton lattices in layered transition metal dichalcogenides, Cr$_x$Ta$_y$S$_z$ (CTS), Cr$_x$Nb$_y$S$_z$ (CNS), and also in the chiral metallic alloy Co$_8$Zn$_{10}$Mn$_2$ above room temperature \cite{winter2026quasi1dspintextureschiral, PhysRevLett.108.107202, PhysRevB.92.220412, https://doi.org/10.1002/apxr.202200116, https://doi.org/10.1002/adma.202101131, Xie, Goodge}. The soliton period in CTS is much smaller than CNS, implying stronger DMI. The Ta-based compounds realize more strongly confined solitons and correspondingly flatter magnon bands, making them a promising platform to investigate the effects discussed below.

\begin{figure}[t]
    \centering
    \includegraphics[width=\linewidth]{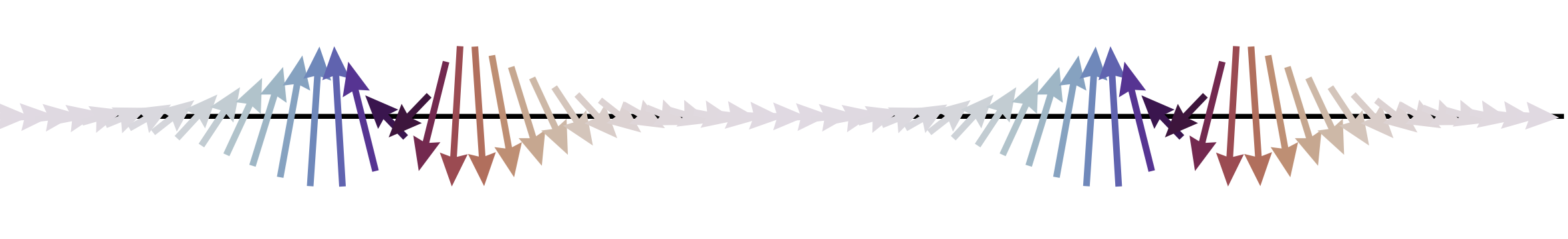}
    \caption{Schematic of the quasi-one-dimensional soliton lattice along the $x$ axis for parameters $J=1$, $D=0.4$, and $B_x=0.0945$. Two unit cells are shown, each containing $N=33$ spins.}
    \label{fig:CSL}
\end{figure}

\textit{Model Hamiltonian.}
We consider a minimal one-dimensional spin chain with isotropic nearest-neighbor Heisenberg interaction (with strength $J$) and DMI (with strength $D$) along the $\hat{z}$ direction. Additionally, a uniform external magnetic field ($\vec{B}_x$) is applied along the $\hat{x}$ direction. The DMI term arises due to the lack of inversion symmetry in the system. The Hamiltonian of our system is given by
\begin{align}
    \mathcal{H} = \sum_{\langle i j \rangle} \left[-J \vec{S}_i \cdot \vec{S}_j-D\hat{z}\cdot (\vec{S}_i \times \vec{S}_j)\right]- \vec{B}_x\cdot \sum_i \vec{S}_i.\label{eq1}
\end{align}
Here, $\vec{S}_i= (S_{i,x}, S_{i,y}, S_{i,z})^T$ with $|\vec{S}_i|=S$ represents the classical spin vector at lattice site $i$, determined by the angles $\theta_i$ and $\phi_i$ of $\vec{S}_i$ with the $z$ and $x$ axes, respectively, in the global spin coordinate system. Below a critical magnetic field $B_c \approx \pi^2 D^2S/16 J$, the ground state of this Hamiltonian is a chiral soliton lattice, see Fig.~\ref{fig:CSL} \cite{ winter2026quasi1dspintextureschiral, PhysRevB.92.220412, PhysRevLett.108.107202, KISHINE20151, PhysRevB.111.024411, hallerQuantumSkyrmionLattices2022, 842q-ghyt, haller2024, PhysRevB.108.054426, KISHINE20151, HANNA1998824, zgqj-tyfd}. Above the critical field, the ground state transitions to a field-polarized phase, where a soliton becomes an excitation. Throughout the paper, we use units where $\hbar=g\mu_B=1$.

\begin{figure*}[t]
  \centering
  \subfloat[\label{fig:sub_a}]{%
    \begin{minipage}{0.48\textwidth}
      \centering
      \includegraphics[width=\linewidth]{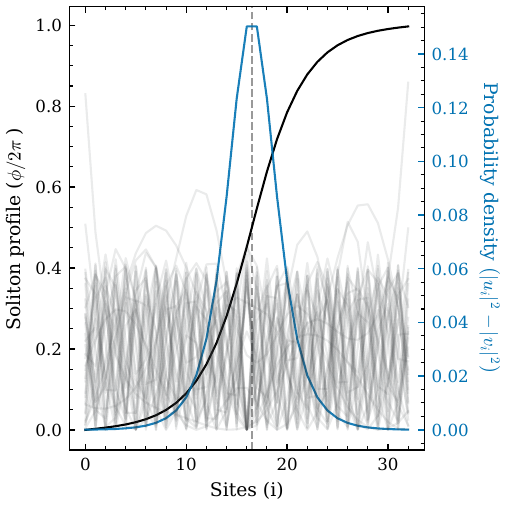}
    \end{minipage}%
  }
  \hfill
  \subfloat[\label{fig:sub_b}]{%
    \begin{minipage}{0.48\textwidth}
      \centering
      \includegraphics[width=\linewidth]{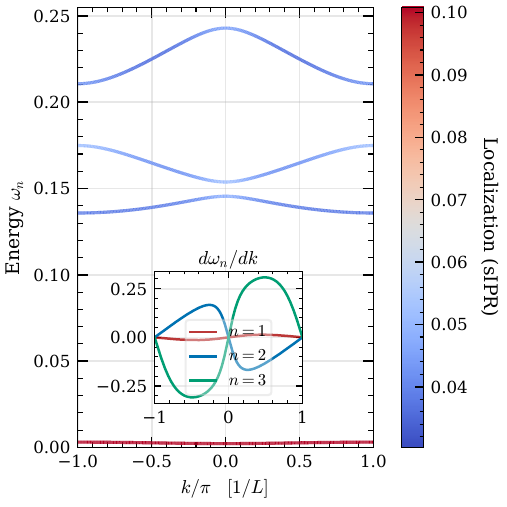}
    \end{minipage}%
  }

  \caption{ (a) Real-space probability density at Bloch momentum $k=0$ (blue line) for positive-energy magnon bands. The black line is the angular dependence $\phi_i$ of the classical spins within one unit cell of the soliton lattice. The vertical dotted line is the center of the soliton. The maximum probability density of the magnon wavefunction is at the soliton center, which corresponds to the lowest flat band of the magnon. Due to a pinning potential, we also have localized modes at the two edges (gray lines) \cite{my_supplement}. (b) Low-energy band structures (scaled with $J$) and symplectic inverse participation ratios (sIPR) of the magnon bands, indicating localization of the magnon wave functions in the lowest band.
  }
  \label{fig:prob_density}
\end{figure*}

\begin{figure*}[t]
  \centering
  \subfloat[\label{fig:qm_a}]{%
    \begin{minipage}{0.49\textwidth}
      \centering
      \includegraphics[width=\linewidth]{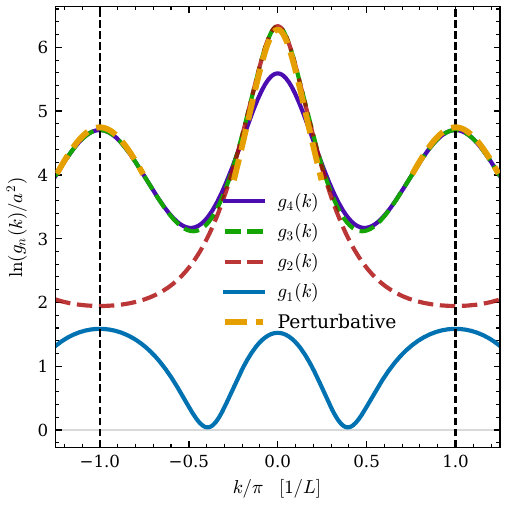}
    \end{minipage}%
  }
  \hfill 
  \subfloat[\label{fig:qm_b}]{%
    \begin{minipage}{0.49\textwidth}
      \centering
      \includegraphics[width=\linewidth]{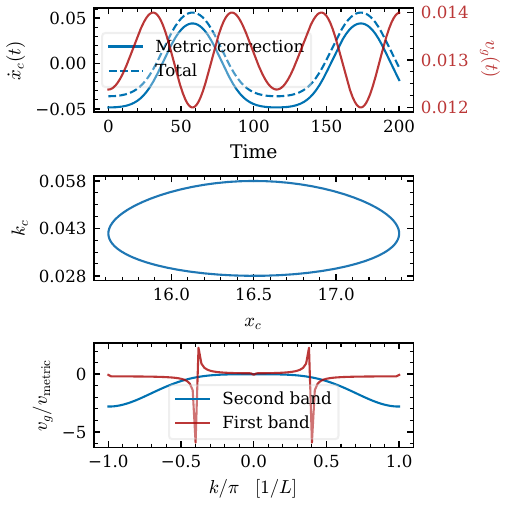}
    \end{minipage}%
  }

  \caption{(a) The quantum metric ($\log$ scale) for the lowest magnon bands. It becomes large near avoided crossings. The yellow curves show analytical fits based on a two-band approximation \cite{PhysRevB.82.245113, krk8-655j, PhysRevB.110.245419, 10.1093/nsr/nwae334}. For the lowest band, the metric cannot be captured by a simple two-band model, and multi-band corrections are required. The lowest flat band is also isolated from other positive energy bands. The particle amplitude $u_i$ dominates over the hole amplitude $v_i$ across the Brillouin zone. The magnitude of the metric for the lowest magnon band is much smaller than the other bands. (b) Center of the wave-packet dynamics as a function of time and phase space trajectory. We can see that the group velocity contribution is negligible compared with the total velocity throughout the time evolution of the wave packet. The metric term flips sign after a certain time and then returns due to the bounded motion within the quantum metric-induced potential. The motion of the center $x_c$ is also bounded around the soliton core.}
  \label{fig:metric}
\end{figure*}

\textit{Magnon excitations.} 
To study the quantum fluctuations of spins on this classical soliton lattice with $S=1$, we perform a Holstein-Primakoff transformation \cite{PhysRev.58.1098} within the linear spin-wave approximation. The classical soliton lattice has a period of $N$ sites, with \(N \propto (4J/\pi D)\,\ln\!\left[16B_c/(B_c-B_x)\right]\) \cite{KISHINE20151, Shinozaki2018} and the Hamiltonian in momentum space is given by
\begin{align}
    \mathcal{H}&= \mathcal{H}_0+\frac{1}{2}\sum_k \Psi_k^\dagger M_k \Psi_k,
\end{align}
with
\begin{align}
    \Psi_k
&=
    \left(
        a_{k,1}, a_{k,2}, \cdots, a_{k, N}, a_{-k,1}^\dagger, a_{-k,2}^\dagger,  \cdots, a_{-k, N}^\dagger
    \right)^T, \label{eq:Psik} \nonumber\\
    M_k
&=
    \begin{pmatrix}
        A_{k} & B_{k}\\
        B_{-k}^* & A_{-k}^T
    \end{pmatrix}.
\end{align}
Here, $a_{k,\alpha}$ denotes the Holstein-Primakoff boson, where $\alpha$ labels the $N$ sublattices in the magnetic unit cell given by the soliton size, and the Bloch momentum $k$ takes values in the first Brillouin zone $[-\pi/L, \pi/L]$, where $L = N a$ and $a$ is the lattice spacing. $\mathcal{H}_0$ is the classical ground state energy. Since the Hamiltonian contains anomalous magnon-pairing terms, a bosonic Bogoliubov-Valatin transformation is necessary to find its eigenmodes, i.e., a transformation matrix $T$ such that $\Psi_k=T \Phi_k$, where $\Phi_k= (b_{k,1}, \cdots, b_{k,N }, b_{-k,1}^\dagger, \cdots, b_{-k, N}^\dagger)^T$. $T$ diagonalizes $M_k$, i.e., $T^\dag M_k T = E_k$ with diagonal matrix $E_k$, as well as preserves bosonic commutation relations, i.e., $T^\dag \Sigma T = \Sigma$ with $\Sigma= \sigma_z\otimes \mathds{1}_{N}$. This means that $T$ must be pseudo-unitary \cite{my_supplement}.

From the Heisenberg equation of motion, one finds that the dynamical matrix $\mathcal{D}_k = \Sigma M_k$ governs the dynamics of the Bogoliubov-Nambu spinors, $d \Psi_k(t)/dt=-i\mathcal{D}_k\Psi_k(t)$. We obtain the magnon spectrum and wavefunctions by diagonalizing the pseudo-Hermitian dynamical matrix. The magnon wave functions $\ket{\psi_{n}(k)}$ are the right eigenvectors of $\mathcal{D}_k$ with energy $\omega_n(k)$, normalized via the Krein product $\bra{\psi_{n}(k)}\Sigma\ket{\psi_{m}(k)}=\text{sgn}[\omega_n(k)] \delta_{mn}$ \cite{krk8-655j}. The wave functions consist of particle and hole sectors as $\psi_n(k)=(u_{nk}, v_{nk})^T$. The symplectic quantum metric of the $n$th band in one dimension is defined as,
\begin{align}
    g_n(k)= \bra{\partial_{k}\psi_n}\Sigma (\mathds{1}_{2N}-\ket{\psi_n}\bra{\psi_n}\Sigma)\ket{\partial_{k}\psi_n}.
\end{align}
As shown in Fig.~\ref{fig:prob_density}(b), the lowest-energy magnon bands are nearly flat, and the group velocity is significantly smaller than that of the other bands. For a given band $n$ and momentum $k$, the symplectic inverse participation ratio (sIPR) is defined as, 
\begin{equation}
    \text{sIPR}_n(k) = \frac{\sum_i 
    \left(|u_i|^2-|v_i|^2\right)^2}{\left(\sum_i (|u_i|^2-|v_i|^2)
    \right)^2}
\end{equation}
where $i \in \{1, \ldots, N\}$ labels the components of the vectors $u_{nk}$ and $v_{nk}$. The lower bound of the sIPR is given by $1/N$ for a delocalized wavefunction (where $u_i, v_i$ are independent of $i$), whereas it reaches $1$ for a perfectly localized wavefunction. For the lowest band, we observe an enhanced sIPR relative to the delocalized bound for magnons. A semiclassical LLG study also reported localization of the lowest frequency resonance peak around the soliton core \cite{xbjs-pfhg}. Due to the strong localization of the magnons and very small group velocity, conventional group-velocity-mediated transport is strongly suppressed. By harnessing the finite symplectic quantum metric contribution of the nearly flat band, see Fig.~\ref{fig:metric}(a), we can control and modify the wave-packet dynamics using a weak external perturbation \cite{my_supplement}.

\textit{Wave-packet dynamics.}
We construct a wave packet in the $n$th nondegenerate magnon BdG band as \cite{PhysRevB.59.14915, RevModPhys.82.1959} 
\begin{align}
  \ket{W(t)} &= \int_{-\pi/L}^{\pi/L} \frac{dk}{2\pi}\,\tw(k,t)\,e^{ikx}\,\ket{\psi_{n}(k)}.
\end{align}
where $\tw(k,t) = |\tw(k,t)|\,e^{i\gamma(k,t)}$ is a smooth complex envelope
peaked at Bloch momentum $\kc(t)$ \cite{PhysRevB.99.121111, PhysRevLett.122.227402}. Moreover, $\sigma_{\text{env}}^2=(2\pi)^{-1}\int dk(\partial_k|\tw|)^2$ is the spread of the wave packet spanning several magnetic unit cells. For a dispersion with small $\partial^2\omega_n(k)/\partial k^2$, we can apply a rigid-envelope approximation in which the spread of the wave packet is time-independent \cite{my_supplement}.

To induce dynamics in the system, we apply an effective, weakly position-dependent chemical potential to the magnons. Then, the perturbation Hamiltonian can be expressed as
\begin{equation}
  \delta \mathcal{H} =\frac{1}{2}\sum_k \Phi_k^\dag \mu(\hat{x}) \Phi_k,
\end{equation}
where the diagonal elements of the matrix $\mu(x)$ contains the chemical potential for different bands.
Thus, the perturbation term due to the chemical potential for $n$th magnon band can be written as,
\begin{equation}
    \delta \mathcal{V}_n= \frac{1}{2} \sum_k \mu_n(\hat{x})( b_{k,n}^\dagger b_{k,n}+ b_{-k,n}^\dagger b_{-k,n})
\end{equation}
where $\hat{x}=i\partial_k+\mathcal{A}_n(k)$ with the symplectic Berry connection $\mathcal{A}_n(k)$ \cite{my_supplement}. Such a chemical potential changes the local magnon density in different bands. After Taylor expanding $\mu(\hat{x})$ up to second order in $(\hat{x}-x_c)$, and applying the first-order perturbation correction, the dynamics of the center of the wave packet are given by,
\begin{align}
  \dot{x}_c
  &= \partial_{k_c}\omega_n(\kc)
   + \frac{1}{2}\,\partial_{k_c}\gnn(\kc)\,\partial^2_{x_c} \mu_n(\xc),
  \label{eq:xc_eom1}\\
  \dot{k}_c
  &= -\partial_{x_c} \mu_n(\xc)
    - \frac{1}{2}\,\left[\gnn(\kc)+\sigma_{\text{env}}^2\right]\,\partial^3_{x_c} \mu_n(\xc). \label{eq:kc_eom1}
\end{align}
We assumed that during the evolution of the wave packet interband transitions due to Landau-Zener (LZ) tunneling are suppressed, so the single-band semiclassical description remains valid throughout the evolution. 
According to Eq.~\eqref{eq:xc_eom1}, the center of the wave packet in real space is affected by the quantum metric term. Near the flat-band limit, the group velocity term $v_g \approx \partial \omega_n/\partial k_c$ is small, and the dynamics will be mainly governed by the quantum metric. From Eq.~\eqref{eq:kc_eom1}, one finds that the force on the wave packet is also modified by the metric term.

To better understand the dynamics of the center of the wave packet, we expand the perturbing chemical potential around the soliton center $x_0$. Choosing the potential to have a maximum at the center of the soliton, i.e., $\partial \mu_n/\partial x_0=0$ and $\partial^2 \mu_n/\partial x_0^2 < 0$, we obtain,
\begin{align}
    \mu_n(x_c) &\approx \mu_n(x_0)-\frac{\mu_n^0(x_c-x_0)^2}{2},\\ \mu_n^0\frac{a^2}{J} &=\frac{a^2}{J}\abs{\frac{\partial^2 \mu_n}{\partial x_0^2}} \ll 1.
\end{align}
For the dynamics, we focus on the lowest magnon band, which is nearly flat and more populated than other bands at low temperatures due to the bosonic nature of the magnons \cite{PhysRevResearch.1.032023, giesen2026resonantexcitationslowfrequency}. We choose the initial wave packet center to be at the soliton center $(x_0)$ and choose the initial momentum as $k_I=k_0+\delta$, where $g(k_0)$ is minimal with $\partial g(k_0)/ \partial k_0=0$.

The approximate trajectory of the center of the magnon wave packet is then given by \cite{my_supplement},
\begin{align}
    x_c(t) &=x_0-\frac{\Omega \delta}{\mu_n^0}\sin(\Omega t), & 
    k_c(t) &=k_0+\delta\cos(\Omega t).
\end{align}
If the motion of the wave packet is confined around the local minimum $k_0$ of the quantum metric, the angular frequency is given by $\Omega= \mu_n^0\sqrt{g''(k_0)/2}$ \cite{my_supplement}. The trajectory in phase space satisfies the ellipse equation,
\begin{align}
    \frac{(x_c(t)-x_0)^2}{c_x^2}+\frac{(k_c(t)-k_0)^2}{c_k^2}=1,
\end{align}
with semi-axes $c_x= \Omega \delta/\mu_n^0$ and $c_k=\delta$. Hence, the magnon undergoes periodic motion covering a phase space area $\pi \delta^2 \sqrt{g''(k_0)/2}$, independently of the external perturbation strength. It depends only on the initial offset momentum of the wave packet center and the second derivative of the intrinsic symplectic quantum metric at the minimum in the momentum space.
In the flat-band limit, we also have
\begin{align}
    \frac{d^2k_c}{dt^2}+\frac{(\mu_n^0)^2}{2}\frac{\partial g(k_c)}{\partial k_c}=0.
\end{align}
Comparing the above equation with the force equation in real space, one can interpret the symplectic quantum metric as generating an effective potential $\propto (\mu_n^0)^2g(k_c)/2$ in momentum space. This potential is intrinsic to the band geometry and is scaled by the perturbation strength $\mu_n^0$. 
For our model, assuming that $\mu_n^0=9.45\times 10^{-4} J/a^2$ and $g''(k_0) \approx 9\times 10^3 a^4$, the frequency scale of the wave packet is approximately $\Omega \approx 0.06 J$, significantly smaller than the band gap between the lowest energy and the first excited state bands. The time period can be estimated as $T \approx 100 /J$.

\textit{Conclusion.}
We have shown that in a chiral soliton lattice, the symplectic quantum geometry of bosonic Bogoliubov bands strongly affects the longitudinal motion of magnon wave packets. The nonlinear spin texture acts as a periodic confinement potential, producing low-energy modes localized around the soliton cores. The localized modes give rise to nearly flat magnon bands in which the conventional group velocity is strongly suppressed. The corresponding symplectic quantum metric remains finite and develops a pronounced momentum dependence around the band gap. Our results provide a direct dynamical interpretation of the symplectic quantum metric: beyond measuring distances between Bogoliubov eigenstates, it acts as an effective momentum-space potential for the adiabatic motion of a bosonic wave packet. Chiral soliton lattices, therefore, offer a pathway for isolating quantum-metric effects in a regime without an intrinsic transverse Berry-curvature response, opening a broader direction for geometry-controlled dynamics in textured magnets and bosonic Bogoliubov systems.

Regarding experimental realizations, a slowly varying magnon chemical potential can be achieved by inhomogeneous spin current injection or local microwave pumping \cite{Levati2025, Demidov2017, Schlitz2025, PhysRevB.94.014412, Wittrock2026, doi:10.1126/science.aak9611, PhysRevX.10.021029}. Recent developments in X-ray momentum microscopy \cite{Wittrock2026} and plasmon-enhanced Brillouin light scattering \cite{PhysRevB.110.184419} could be used to track the magnon wave-packet dynamics in nanoscale devices. 

\textit{Acknowledgments.} The authors are grateful to Vladyslav M.~Kuchkin and Maxime Jamotte for valuable feedback and discussions.
SC acknowledges funding from the Luxembourg National Research Fund under Grant No.~AFR/2000698 for the project ``Topology, Transport and Localization of Magnons in Nonlinear Spin Textures.''

\bibliography{MagnonGeometry}

\clearpage

\onecolumngrid

\setcounter{page}{1}
\renewcommand{\thepage}{S\arabic{page}}

\setcounter{figure}{0}
\renewcommand{\thefigure}{S\arabic{figure}}
\setcounter{table}{0}
\renewcommand{\thetable}{S\arabic{table}}

\newcounter{supplsection}
\renewcommand{\thesection}{S\arabic{supplsection}}
\renewcommand{\theequation}{\thesection.\arabic{equation}}

\newcommand{\suppsection}[1]{%
  \refstepcounter{supplsection}%
  \setcounter{equation}{0}%
  \section*{\texorpdfstring{\thesection\quad #1}{#1}}%
}

\begin{center}
{\Large\bf Supplemental Material}\\[0.5cm]
for\\[0.3cm]
{\bf Magnon dynamics driven by the symplectic quantum metric in a chiral soliton lattice}
\end{center}

\suppsection{Continuum Limit}

In this section, we will explain how to construct the continuum limit of the magnetic system. To be specific, we focus on the chiral soliton lattice phase with a periodic classical ground state. We retain only one unit cell consisting of $N$ sites and can rewrite the Hamiltonian as,
\begin{align}
    \mathcal{H}
&= 
    \sum_{i=1}^N \left[-J \vec{S}_i\cdot \vec{S}_{i+1}- D \hat{z}\cdot (\vec{S}_i \times \vec{S}_{i+1})-\vec{B}\cdot \vec{S}_i \right], \quad  \
    \vec{B}=\begin{pmatrix}
        B_x, 0, B_z
    \end{pmatrix}^T. \label{app:eqhamil}
\end{align}
Each of the $N$ spins has $3$ degrees of freedom $S_{x,y,z}$. As the magnitude of each spin vector is constant, $\sqrt{S_x^2 + S_y^2 + S_z^2} = S$, the total number of degrees of freedom reduces to $2N$. We can avoid solving the resulting $2N$ nonlinear coupled algebraic equations by going to the continuum limit, where the resulting two coupled nonlinear differential equations can be solved using analytic or numerical techniques.

We take the continuum limit by assuming that the spins change slowly over the lattice spacing $a$, where $a$ is very small compared to the length $L$ of the unit cell, i.e., $N = L/a \gg 1$. We can then replace the discrete set of spins $\vec{S}_i$ by a continuous vector field $\vec{S}(x)$ at each position $x = i a$. Taylor expanding the spin vector field near $x$ up to the second order in $a$,
\begin{align}
    \vec{S}(x+a) &\approx \vec{S}(x) + a \partial_x \vec{S}(x)+\frac{a^2}{2}\partial_x^2\vec{S}(x)
\end{align}
allows us to write the exchange term as,
\begin{align}
    \vec{S}_i\cdot \vec{S}_{i+1} \longrightarrow \vec{S}(x)\cdot \vec{S}(x+a)
    &\approx \vec{S}(x)\cdot \left[\vec{S}(x)+a\partial_x \vec{S}(x)+\frac{a^2}{2} \partial_x^2\vec{S}(x)\right]\notag \\
    &\approx S^2+\frac{a}{2}\frac{\partial}{\partial x} S^2+ \frac{a^2}{2} \vec{S}(x)\cdot \frac{\partial^2 \vec{S}(x)}{\partial x^2}
\end{align}
Since $\vec{S}(x)$ has a fixed length we have $\vec{S}(x) \cdot \vec{S}(x) = S^2$. Taking derivatives of this equation leads to
\begin{align}
    \vec{S}(x)\cdot \partial_x \vec{S}(x)&=0\\
    \vec{S}\cdot \partial_x^2 \vec{S}(x)&= -(\partial_x \vec{S}(x))^2
\end{align}
Therefore, the Heisenberg exchange term becomes,
\begin{align}
    \mathcal{H}_J&= -J
    \left(S^2-\frac{a^2}{2}(\partial_x \vec{S}(x))^2\right)
\end{align}
For the DMI term, we take the continuum limit analogously,
\begin{align}
    \vec{S}_i\times \vec{S}_{i+1}&\longrightarrow \vec{S}(x)\times \left[\vec{S}(x)+a\partial_x \vec{S}(x)+\frac{a^2}{2} \partial_x^2\vec{S}(x)\right]
\end{align}
Here, the leading contribution emerges to the first order in the gradient term,
\begin{align}
    \mathcal{H}_{\text{DMI}}&= -Da\hat{z}\cdot\left(\vec{S}(x)\times \partial_x \vec{S}(x)\right)
\end{align}
Taking the continuum limit implies changing the summation to an integral,
\begin{align}
\sum_{i=1}^N&\longrightarrow \frac{1}{a} \int_{0}^{L} dx
\end{align}
Hence, the Hamiltonian becomes
\begin{align}
\mathcal{H}&= \frac{1}{a} \int_{0}^{L}dx \left\{-J\left[S^2-\frac{a^2}{2}(\partial_x \vec{S}(x))^2\right]-Da \hat{z}\cdot\left[\vec{S}(x)\times\partial_x \vec{S}(x)\right]-\vec{B}\cdot\vec{S}(x)\right\} 
\end{align}
We express the spin vector field in terms of polar and azimuthal angles $\phi(x)$ and $\theta(x)$ as 
\begin{align}
\vec{S}(x)= S \begin{pmatrix}\sin(\theta)\cos(\phi) \\ \sin(\theta)\sin(\phi) \\ \cos(\theta) \end{pmatrix},
\end{align}
which leads to 
\begin{align}
    \mathcal{H}
&= 
    -NJS^2 + \frac{JS^2a}{2}\int_{0}^{L} dx \left( \theta_x^2+\phi_x^2 \sin^2 \theta \right) -DS^2\int_{0}^{L} dx \phi_x \sin^2 \theta -\frac{S}{a} \int_{0}^{L} dx(B_x\sin{\theta}\cos{\phi}+B_z\cos{\theta}) \notag \\ 
&= 
    -NJS^2+ \int_{0}^{L}dx~ \mathcal{E}(x) 
\end{align}
where we used the notation $\theta_x \equiv \partial_x \theta$ and $\phi_x \equiv \partial_x \phi$ and assumed $\vec{B} = (B_x, 0, B_z)^T$. In the last line, we introduced the energy density,
\begin{gather}
    \mathcal{E}(x) = \frac{JS^2a}{2}(\theta_x^2+\phi_x^2 \sin^2 \theta )- DS^2\phi_x\sin^2 \theta- \frac{B_zS}{a}\cos \theta -\frac{B_xS}{a}\sin \theta \cos \phi \label{eqEnergy}
\end{gather}
To determine the ground state of this Hamiltonian, we solve the Euler-Lagrange equations for the fields $\phi$ and $\theta$. The differential equation for $\phi$ reads,
\begin{align}
\frac{\partial\mathcal{E}}{\partial\phi} &= \frac{\partial}{\partial x}\left[\frac{\partial\mathcal{E}}{\partial \phi_x}\right] 
& \Longleftrightarrow & & \frac{\partial^2\phi}{\partial x^2}&=\frac{B_x}{JSa^2}\frac{\sin \phi}{\sin \theta }-\frac{2\cos \theta }{\sin \theta }\frac{\partial\theta}{\partial x}\left(\frac{\partial\phi}{\partial x}-\frac{D}{Ja}\right)\label{eqphi2}
\end{align}
Analogously, the Euler-Lagrange equation for $\theta$ is given by
\begin{align}
    \frac{\partial\mathcal{E}}{\partial\theta} &= \frac{\partial}{\partial x}\left[\frac{\partial\mathcal{E}}{\partial \theta_x}\right]
& \Longleftrightarrow & & 
    \frac{\partial^2\theta}{\partial x^2} &=\frac{1}{2}\sin(2\theta)\left(\frac{\partial \phi}{\partial x}\right)^2-\frac{D}{Ja}\sin(2\theta)\frac{\partial\phi}{\partial x}+\frac{B_z}{JSa^2}\sin(\theta)-\frac{B_x}{JSa^2}\cos(\theta)\cos(\phi) \label{eqtheta2}
\end{align}
Equations~\eqref{eqtheta2} and \eqref{eqphi2} form a set of two coupled nonlinear differential equations, which in general need to be solved numerically.

\subsection{Spins confined in the \texorpdfstring{$x-y$}{x-y} plane}

An analytical solution can be obtained for the case when $B_z = 0$, as in this case a constant $\theta(x)= \pi/2$ is a solution of the Euler-Lagrange equation, i.e., the spins remain confined in the $x-y$ plane. The in-plane state of the spins is then governed by $J$, $D$, and $B_x$. The differential equation \eqref{eqphi2} reduces to,
\begin{align}
    \frac{\partial^2\phi}{\partial x^2}&=\frac{B_x}{JSa^2}\sin \phi \equiv m^2\sin \phi \label{sine-gordon}
\end{align}
which is exactly the famous static sine-Gordon equation with effective mass $m=\sqrt{B_x/JSa^2}$ for the case $B_x > 0$. The solution for $B_x < 0$ corresponds to shifting $\phi(x) \to \phi(x) + \pi$. In the absence of the magnetic field ($B_x=m=0$), the ground state is a spin spiral,
\begin{align}
    \phi(x)&=\phi_0+q_0x,\quad q_0=\frac{D}{Ja}
\end{align}
In contrast, for $B_x \neq 0$, the solution of the $\phi(x)$ has the form of a soliton,
\begin{align}
     \phi(x)= 4\tau\arctan(e^{m(x-x_0)})\label{phi}
\end{align}
where $x_0$ is a free parameter denoting the center of the soliton. The so-called kink ($\tau=1$) or anti-kink ($\tau=-1$) solution interpolates from $\phi = 0$ at $x \to -\infty$ to $\phi = 2\pi$ ($\phi = -2\pi$) at $x \to \infty$. For $B_x \neq 0$, we thus generally obtain a nonlinear spin texture. For a small but finite magnetic field, the spin spiral ground state evolves into a soliton state. 

Since the kink and anti-kink solutions approach their asymptotic values $\phi = 0, \pm 2\pi$, where the energy density becomes minimal, exponentially fast as a function of $|x - x_0|$, we can construct approximate solutions by concatenating solitons. As a result of remnant soliton interactions, this results in a soliton lattice.

The chiral soliton lattice phase thus emerges from the competition between the DMI, which favors a spin spiral in the $x-y$ plane, and a magnetic field perpendicular to the DMI vector, which favors a polarized state along the $x$ axis.

\begin{figure}[t]
  \centering

  \subfloat[]{%
    \label{fig:discrete_a}%
    \begin{minipage}[t]{0.49\linewidth}
      \centering
      \includegraphics[width=\linewidth]{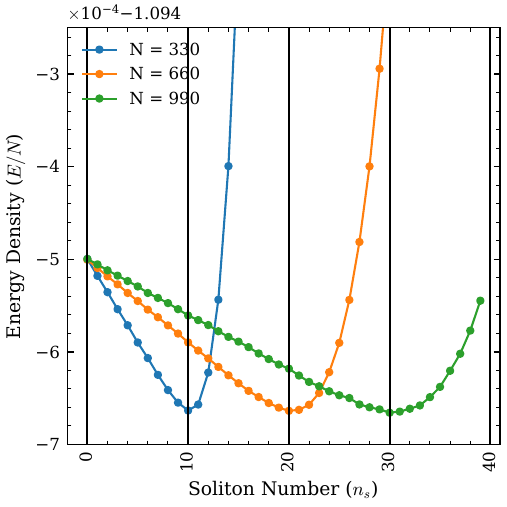}
    \end{minipage}%
  }
  \hfill
  \subfloat[]{%
    \label{fig:discrete_d}%
    \begin{minipage}[t]{0.49\linewidth}
      \centering
      \includegraphics[width=\linewidth]{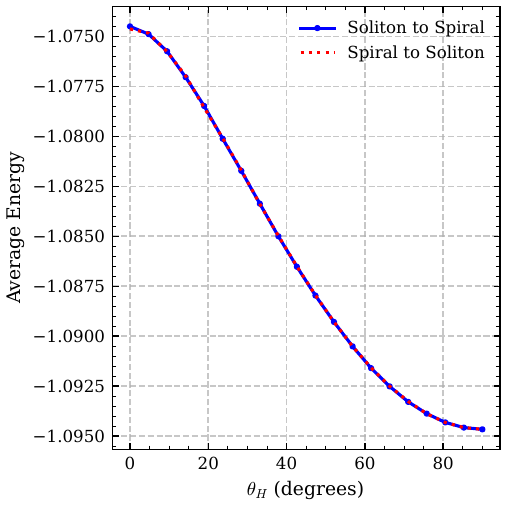}
    \end{minipage}%
  }

  \caption{(a) Classical ground-state energy minimization from different winding-number initial states using the gradient descent algorithm. We can see that the periodicity of the soliton unit cell remains invariant under a change of the system size. (b) We have added a position-dependent potential term in our Hamiltonian, which can act as a perturbed potential for the spin spiral phase and also as a constant external field along $x$ for the soliton phase, in different limits. The modified angle $\theta_H$ with the $z$ axis for the position dependent potential is given by, $\theta_H(\tau)=(1-\tau)\phi_{\text{spiral}}+\tau \phi_{\text{soliton}}$. When $\tau=0$, we have a perturbed spiral; when $\tau=1$, we have a soliton. We have performed the Monte Carlo simulations for each $\tau$ and connected the spiral and soliton phases continuously. The energy varies smoothly with $\tau$ between these two phases, indicating a crossover rather than a phase transition \cite{https://doi.org/10.1002/adma.202101131, WANG2025108336, PhysRevB.87.104403, PhysRevB.101.184425, PhysRevB.110.144426, Clements2017,doi:10.7566/JPSJ.85.013707, PhysRevB.96.184423}.}
  \label{fig:discrete_limit_supp}
\end{figure}

 \suppsection{Numerical energy minimization for the discrete system}
Now, we consider $n_s$ stable solitons in our system of size $L_s$, and the soliton unit cell length is commensurate with the system size. Consider that there are $N$ spins within the single soliton with the periodicity of one soliton $L$. For the simplicity we take, $\frac{L}{N}=a=1$, where $a$ is the lattice constant. Thus, the spins are periodic with the periodicity length $L$. In the momentum space, the Brillouin zone is reduced to $\left[-\pi/N, \pi/N\right]$. To minimize the energy in the discrete spin system, instead of solving $2N$ coupled algebraic equations simultaneously, we use Metropolis with Monte Carlo sampling or the gradient descent algorithm to find the ground state of our Hamiltonian. The results of these simulations and the crossover from the soliton to the spiral phase are shown and discussed in Fig.~\ref{fig:discrete_limit_supp}.

\suppsection{Magnon Hamiltonian and bosonic Bogoliubov transformation}

In this section, we show how to derive the magnon theory in the linear spin-wave approximation for the chiral spin lattice phase. We start by writing the classical spin Hamiltonian~\eqref{app:eqhamil} as,
\begin{align}
    \mathcal{H}
&= 
    \sum_{\langle i j \rangle} \vec{S}_i^T M \vec{S}_j - \vec{B}\cdot\sum_i \vec{S}_i, 
&
    M
&=
    \begin{pmatrix}
        -J & -D & 0\\
        D & -J & 0\\
        0 & 0 & -J
    \end{pmatrix}, 
    &
    \vec{S}
&=
    \begin{pmatrix}
        S_x \\ S_y \\ S_z
    \end{pmatrix}
\end{align}
We would like to consider the magnon excitations on top of the ground state of this spin Hamiltonian. To apply the Holstein-Primakoff transformation within the linear spin-wave approximation, we first need to perform a local rotation of the spin axis in this Hamiltonian such that in the new basis, the ground state consists of spins aligned along the $z$ axis. In general, this can be achieved using a site-dependent rotation $R_i$ of the spin axis,
\begin{align}
    \vec{S}'_i &= R_i(\theta_i, \phi_i) \vec{S}_i\\
    \vec{S}_i &= [R_i(\theta_i, \phi_i)]^{-1}\vec{S}'_i= [R_i(\theta_i, \phi_i)]^T\vec{S}'_i
\end{align}
where each of the rotation matrices $R_i \in \mathrm{SO}(3)$ is comprised of rotations about the $y$ and $z$ axes, and thus parametrized by two angles $\theta_i$ and $\phi_i$,
\begin{align}
    R^z(-\phi)
&=
    \begin{pmatrix}
        \cos(\phi) & \sin(\phi) & 0\\
        -\sin(\phi) & \cos(\phi) & 0\\
        0 & 0 & 1
    \end{pmatrix}, \notag \\
    R^y(-\theta)
&= 
    \begin{pmatrix}
        \cos(\theta) & 0 & -\sin(\theta)\\
        0 & 1 & 0\\
        \sin(\theta) & 0 & \cos(\theta)
    \end{pmatrix}\notag \\
    R(\theta, \phi)
&=
    R^y(-\theta)R^z(-\phi)=\begin{pmatrix}
        \cos(\theta)\cos(\phi) & \cos(\theta)\sin(\phi) & -\sin(\theta)\\
        -\sin(\phi) & \cos(\phi) & 0\\
        \sin(\theta)\cos(\phi) & \sin(\theta)\sin(\phi) & \cos(\theta)
    \end{pmatrix}
\end{align}
In the new coordinate system, the Hamiltonian can be rewritten as,
\begin{align}
    \mathcal{H}
&=
    \sum_{\langle i j\rangle} \vec{S}'^T_i (R_i M R_j^T) \vec{S}'_j- \vec{B}\cdot\sum_i R_i^T \vec{S}'_i 
= 
    \sum_{\langle i j\rangle} \vec{S}'^T_i \tilde{M}_{ij} \vec{S}'_j- \sum_i \tilde{B}_i \cdot \vec{S}'_i
\end{align}
with rotated coupling matrix and magnetic field,
\begin{align}
    \tilde{M}_{ij}
&= 
    R_i M R_j^T=
    \begin{pmatrix}
    \tilde{M}_{xx} & \tilde{M}_{xy} & \tilde{M}_{xz}\\
    \tilde{M}_{yx} & \tilde{M}_{yy} & \tilde{M}_{yz}\\
    \tilde{M}_{zx} & \tilde{M}_{zy} & \tilde{M}_{zz}
    \end{pmatrix} \notag \\
    \tilde{B}_i
&= 
    R_i \vec{B}
\end{align}
where we have suppressed site indices $i,j$ on the individual matrix elements of $\tilde{M}_{ij}$.

The aim is now to determine the rotation matrices $R_i$ in such a way that in the rotated basis, the ground state has all spins aligned along the $z$ axis, i.e., $\vec{S}'_i = (0,0,S)$ for all $i$. 
Once such rotation matrices have been found, we define the Holstein-Primakoff transformation on the aligned spins within the linear spin-wave approximation as follows,
\begin{align}
S'^{x}_i&=\sqrt{\frac{S}{2}}(a_i+a_i^\dagger), &
S'^{y}_i&= -i\sqrt{\frac{S}{2}}(a_i-a_i^\dagger), & 
S'^z_i&= S- a_i^\dagger a_i
\end{align}
where $a_i$ denote bosonic operators. We proceed by retaining terms up to the second order in the bosonic operators.

Terms in the Hamiltonian containing ${S'}^x{S'}^z$, ${S'}^y {S'}^z$, ${S'}^x$ or ${S'}^y$, which might emerge due to the rotation, would give rise to linear terms in the bosonic Hamiltonian. 
However, when performing the linear spin wave approximation about the ground state, the coefficients of the linear operators vanish because the ground state is stable.

Hence, the Hamiltonian is composed of quadratic terms in the $a_i$ operators, notably local terms $\propto a_i^\dag a_i$, nearest-neighbor hopping terms of the form $a^\dag_i a_{i+1}$, and nearest-neighbor pairing terms $\propto a_i a_{i+1}$,

\begin{align}
    \mathcal{H} 
&= 
    \mathcal{H}_0+\sum_{i=1}^N \left\{ \frac{S}{2}\begin{pmatrix}
        a_i & a_i^\dagger
    \end{pmatrix}\begin{pmatrix}
        1 & -i\\
        1 & i
    \end{pmatrix}\begin{pmatrix}
        \tilde{M}^{i,i+1}_{xx} & \tilde{M}^{i,i+1}_{xy}\\
        \tilde{M}^{i,i+1}_{yx} & \tilde{M}^{i,i+1}_{yy}
    \end{pmatrix}\begin{pmatrix}
        1 & 1\\
        -i & i
    \end{pmatrix}\begin{pmatrix}
        a_{i+1}\\
        a_{i+1}^\dagger
    \end{pmatrix}-S\tilde{M}^{i,i+1}_{zz} (a_i^\dagger a_i+a_{i+1}^\dagger a_{i+1})+\tilde{B}_{z,i} a_i^\dagger a_i \right\}
    \label{Hbdg}
\end{align} 
where the constant $\mathcal{H}_0$ is the energy of the classical ground state, which we ignore from now on, and we assumed that $\vec{B} = (B_x, 0, 0)^T$. In this case, the effective field is $\tilde{B} = (0,0,\tilde{B}_z)$ with $\tilde{B}_{z,i} = R_{i,31} B_x$.

Next, we Fourier transform the magnon Hamiltonian $\mathcal{H}$ to momentum space, assuming that the system is in the chiral soliton lattice phase. In this case, the classical spin state is periodic with a periodicity of $N$ lattice sites. Therefore, $\phi(x)$ and $\theta(x)$ are periodic functions of $x$ with period $L = N a$, where $a$ is the lattice spacing. The same periodicity holds for the coupling terms $\tilde{M}$ as well as for the effective magnetic field $\tilde{B}_{z,i}$. Therefore, the magnon Hamiltonian can be regarded as a bosonic tight-binding Hamiltonian with a unit cell consisting of $N$ lattice sites. This allows us to perform a Fourier transformation to Bloch momenta $k \in [-\pi/L, \pi/L]$,

Writing $a_i = a_{m N + \alpha} \equiv a_{m,\alpha}$ where $m$ counts the unit cells and $\alpha \in [1, \ldots, N]$ is the coordinate within each unit cell, we have

\begin{align}
    a_{m,\alpha}= \frac{1}{\sqrt{N}} \sum_k e^{-ik(m L+ \alpha a)} a_{k, \alpha}
\end{align}
Using this Fourier transformation, we obtain the magnon Bloch Hamiltonian after some manipulations,
\begin{align}
    \mathcal{H}_{\text{BdG}}
&=
    \frac{S}{2}\sum_{\alpha\alpha'} \sum_k \left[\tilde{M}_{xx}^{\alpha\alpha'}-\tilde{M}_{yy}^{\alpha\alpha'}-i(\tilde{M}_{xy}^{\alpha\alpha'}+\tilde{M}_{yx}^{\alpha\alpha'})\right]e^{ika} a_{k,\alpha} a_{-k,\alpha'}+\text{h.c.}\nonumber\\
&+
    \frac{S}{2}\sum_{\alpha\alpha'} \sum_k \left[\tilde{M}_{xx}^{\alpha\alpha'}+\tilde{M}_{yy}^{\alpha\alpha'}+i(\tilde{M}_{xy}^{\alpha\alpha'}-\tilde{M}_{yx}^{\alpha\alpha'})\right ]e^{ika} a_{k,\alpha} a_{k,\alpha'}^\dagger+\text{h.c.}\nonumber\\
&-
    \sum_{\alpha\alpha'} \sum_k \tilde{M}_{zz}^{\alpha\alpha'} S[a_{k,\alpha}^\dagger a_{k,\alpha}+ a_{k,\alpha'}^\dagger a_{k,\alpha'}]
    +\sum_{\alpha\alpha'}\sum_k(\tilde{B}_{z,\alpha}a_{k,\alpha}^\dagger a_{k,\alpha}+\tilde{B}_{z,\alpha'}a_{k,\alpha'}^\dagger a_{k,\alpha'})
\end{align}

Hence, the Hamiltonian can be written as,
\begin{align}
    \mathcal{H}_{\text{BdG}} 
&= 
    \frac{1}{2}\sum_k \Psi_k^\dagger M_k \Psi_k
\end{align}
with the $2N$-component Nambu spinor $\Psi_k$ and the $2N \times 2N$ hermitian matrix $M_k$ given by,
\begin{align}
    \Psi_k 
&=
    \left(
        a_{k,1}, a_{k,2}, \cdots, a_{k, N}, a_{-k,1}^\dagger, a_{-k,2}^\dagger,  \cdots, a_{-k, N}^\dagger
    \right)^T\\
    M_k 
&=
    \begin{pmatrix}
        A_{k} & B_{k}\\
        B_{-k}^* & A_{-k}^T
    \end{pmatrix} \label{M_k},
\end{align}
The entries of the $N \times N$ matrices $A_k$ and $B_k$ are functions of the parameters $J$, $D$, $B_x$, as well as the Bloch momentum $k$.\\
As our Hamiltonian is hermitian, we have $M_k=M_k^\dagger$. So from Eq.~\ref{M_k}, we have, 
\begin{align}
    A_k &=A_k^\dagger & A_{-k}^T&=A^*_{-k}\\
    B_k&=B_{-k}^T & B_{k}^\dagger&=B^*_{-k}
\end{align}
Now, we define an operator $X=\sigma_x \otimes \mathds{1}_{N}$, which will swap the particle and hole basis of our spinor $\Psi$.
The $M_k$ transform under this operator as,
\begin{align}
    X M_k^* X&= \begin{pmatrix}
        A_{-k} & B_{-k}\\
        B_{k}^* & A_k^T
    \end{pmatrix}= M_{-k}\\
    X M_k^* X &=M_{-k} \label{PHCons}
\end{align}
This is the particle-hole constraint of the Hamiltonian.\\
Now, we check how the spinor $\Psi$ transforms under $X$,
\begin{align}
    \Psi_k 
&=
    \left(
        a_{k,1}, a_{k,2}, \cdots, a_{k, N}, a_{-k,1}^\dagger, a_{-k,2}^\dagger,  \cdots, a_{-k, N}^\dagger
    \right)^T. \\
    X\Psi_k &= \left(a_{-k,1}^\dagger, a_{-k,2}^\dagger,  \cdots, a_{-k, N}^\dagger,
        a_{k,1}, a_{k,2}, \cdots, a_{k, N}
    \right)^T= \left[\Psi_{-k}^\dagger\right]^T,\\
    \Psi_{-k}^\dagger &=\Psi_k^T X^T=\Psi_k^T X, \quad X=X^T,\\
    \Psi_k &=X [\Psi_{-k}^\dagger]^T,\\
    \Psi_{k,j} &=\sum_l X_{j,l} \Psi_{-k,l}^\dagger.
\end{align}
Inverting the relation $\Psi_{k,j}=\sum_l X_{j,l}\Psi^\dagger_{-k,l}$ with $X^2=\mathds{1}_{2N}$,
\begin{align}
    \Psi_{-k,l}^\dagger &= \sum_m X_{l,m}\,\Psi_{k,m}. \label{PsiDagger}
\end{align}
Now, we can write the commutation relation between $\Psi_i, \Psi_j$ as,
\begin{align}
    \left[\Psi_{k,i}, \Psi_{k',j}\right] &= \left[\Psi_{k,i}, \sum_l X_{j,l} \Psi_{-k',l}^\dagger\right]\\
    &=\sum_l X_{j,l} \left[\Psi_{k,i}, \Psi_{-k',l}^\dagger\right]\\
    &= \sum_l X_{j,l}\Sigma_{i,l}\delta_{k,-k'}\\
    &= \left[\Sigma X\right]_{ij} \delta_{k, -k'}
\end{align}
\subsection{Bogoliubov transformation}

In the following, we will suppress the momentum index $k$ on the spinor $\Psi_k$ and the Hamiltonian matrix $M_k$. The components $i, j \in \{1, \ldots, 2N\}$ of the spinor with a given momentum satisfy the commutation relations,
\begin{align}\label{eq:app_Psi_comm}
    \left[\Psi_i, \Psi_j^\dagger\right]=(\sigma_z\otimes \mathds{1}_{N})_{ij}=\Sigma_{ij}
\end{align}
where $\Sigma= \sigma_z\otimes \mathds{1}_{N}$ is a diagonal $2N \times 2N$ matrix and satisfies $\Sigma^2 = \mathds{1}_{2N}$. We would like to diagonalize the Hamiltonian matrix $M$ using a transformation $\Psi = T \Phi$ to a bosonic spinor $\Phi = (b_{k,1}, b_{k,2}, \cdots, b_{k,N},  b_{-k,1}^\dagger, b_{-k,2}^\dagger, \cdots, b_{-k,N}^\dagger)^T$, which fulfills commutation relations of the form \eqref{eq:app_Psi_comm}. To ensure this, the transformation $T$ needs to satisfy the two relations,\\
$T$ will diagonalize, $M$
\begin{align}
    T^\dag M T &= E,  
\end{align}
where $E$ is a diagonal $2N \times 2N$ matrix of eigenvalues. And,
$T$ will also fulfill the bosonic commutation relations,
\begin{align}
    \Psi_i &= T_{ij}\Phi_j,\\
    [\Psi_i,\Psi_j^\dagger]
    &= [T_{ik}\Phi_k,\Phi_{k'}^\dagger T_{jk'}^*]\\
    &= T_{ik}[\Phi_k,\Phi_{k'}^\dagger]T_{jk'}^*\\
    &= T_{ik}\Sigma_{kk'}T_{jk'}^*.
\end{align}

Since
\begin{equation}
[\Psi_i,\Psi_j^\dagger]=\Sigma_{ij},
\end{equation}
we have
\begin{equation}
T_{ik}\Sigma_{kk'}T_{jk'}^*=\Sigma_{ij}
\implies
T\Sigma T^\dagger=\Sigma
\end{equation}
A matrix $T$ with these properties is called pseudo-unitary, and it fulfills $T^{-1} = \Sigma T^\dag \Sigma$.
\\
The Heisenberg equation of motion for the $i$-th component of the Nambu spinor can be written as
($\mathcal{H}_0$ is constant, so the commutation is zero),
\begin{align}
    i\,\frac{d\Psi_{k,i}}{dt}
    &= \left[\Psi_{k,i},\mathcal{H}\right]
     = \frac{1}{2}\sum_{k'}\sum_{j,l}
       \left[\Psi_{k,i},\,\Psi^\dagger_{k',j}\left(M_{k'}\right)_{jl}\Psi_{k',l}\right]\\
    &= \frac{1}{2}\sum_{k'}\sum_{j,l}\left(M_{k'}\right)_{jl}
       \left(
        \underbrace{\left[\Psi_{k,i},\Psi^\dagger_{k',j}\right]\Psi_{k',l}}_{\text{normal}}
        +\underbrace{\Psi^\dagger_{k',j}\left[\Psi_{k,i},\Psi_{k',l}\right]}_{\text{anomalous}}
       \right).
\end{align}
The normal term contributes when $k=k'$
\begin{align}
    \frac{1}{2}\sum_{k'}\sum_{j,l}\left(M_{k'}\right)_{jl}\Sigma_{ij}\delta_{k,k'}\Psi_{k',l}
    &= \frac{1}{2}\sum_{j,l}\Sigma_{ij}\left(M_{k}\right)_{jl}\Psi_{k,l}
     = \frac{1}{2}\left[\Sigma M_k \Psi_k\right]_i .
\end{align}
The anomalous term is non-vanishing when $k'=-k$;
\begin{align}
    \frac{1}{2}\sum_{k'}\sum_{j,l}\Psi^\dagger_{k',j}\left(M_{k'}\right)_{jl}
        \left[\Sigma X\right]_{il}\delta_{k,-k'}
    &= \frac{1}{2}\sum_{j,l}\left[\Sigma X\right]_{il}
        \left(M_{-k}\right)_{jl}\Psi^\dagger_{-k,j}\\
    &= \frac{1}{2}\sum_{j,l,m}\left[\Sigma X\right]_{il}
        \left(M^T_{-k}\right)_{lj} X_{j,m}\Psi_{k,m}
        \qquad \text{[using Eq.~\eqref{PsiDagger}]}\\
    &= \frac{1}{2}\left[\Sigma X M^{T}_{-k} X\Psi_k\right]_i
     = \frac{1}{2}\left[\Sigma X M^{*}_{-k} X\Psi_k\right]_i\\
    &= \frac{1}{2}\left[\Sigma M_{k}\Psi_k\right]_i ,
\end{align}
where $M^T_{-k}=M^*_{-k}$ follows from $M_{-k}=M^\dagger_{-k}$, and the last line is the
particle-hole constraint of Eq.~\eqref{PHCons} evaluated at $k\rightarrow -k$,
i.e. $X M^*_{-k} X = M_{k}$.
Both terms are therefore equal, and the factor $1/2$ of the Nambu doubling is canceled,
\begin{align}
    i\,\frac{d\Psi_{k,i}}{dt}
    &= \frac{1}{2}\left[\Sigma M_k\Psi_k\right]_i
      +\frac{1}{2}\left[\Sigma M_k\Psi_k\right]_i
     = \left[\mathcal{D}_k\Psi_k\right]_i ,
\end{align}
so that, in matrix form,
\begin{align}
    \frac{d\Psi_k(t)}{dt}=-i\mathcal{D}_k\Psi_k(t),
    \qquad \mathcal{D}_k=\Sigma M_k.
\end{align}
We define the dynamical matrix as, $\mathcal{D}_k=\Sigma M_k$. Under this symmetry operation $X$, the dynamical matrix transforms as,
\begin{align}
    X \mathcal{D}_{-k}^* X= X \Sigma M^*_{-k}  X
\end{align}
Using, $X \Sigma= -\Sigma X$, we have
\begin{align}
    X \mathcal{D}_{-k}^* X&= -\Sigma X M^*_{-k} X\\
    &= -\Sigma M_{k}\\
    &= -\mathcal{D}_{k}
\end{align}
$X \mathcal{D}_{-k}^* X=-\mathcal{D}_k$ represents the particle hole constraint of the bosonic Bogoliubov system \cite{PhysRevB.100.075414, 1sz1-nwzf}.
The dynamical matrix is pseudo-Hermitian, i.e., $\Sigma \mathcal{D}_k^\dag \Sigma = \mathcal{D}_k$ and has real eigenvalues if $M_k$ is positive definite. We assume that the latter is the case in order to ensure a stable ground state. The normal-mode ansatz $\Psi_k(t)=\psi_n(k)\,e^{-i\omega_n(k)t}$ then gives the
time-independent eigenvalue problem
\begin{align}
    \mathcal{D}_k\ket{\psi_n(k)}=\omega_n(k)\ket{\psi_n(k)}.
\end{align}
In terms of the spinor $\Phi$, we can write the relation for the dynamical matrix as,
\begin{align}
    \Phi_{nk}^\dagger \Sigma \left(T_k^{-1} \mathcal{D}_k T_k\right)\Phi_{mk}= \text{sgn}[\omega_n(k)] \omega_{nk}\delta_{nm} b_{nk}^\dagger b_{mk}
\end{align}
We define $\psi_n(k)$ as the $n$th column of the transformation matrix $T$ \cite{krk8-655j}, where $n$ is the band index. $\psi_n(k)$ is the eigenvector of the dynamical matrix and normalized via Krein product \cite{krk8-655j}, $\bra{\psi_{n}(k)}\Sigma\ket{\psi_{m}(k)}=\text{sgn}[\omega_n(k)] \delta_{mn}$, $\text{sgn}[\omega_n(k)]=+1$ for stable positive energy magnon bands.\\
For $N$ orbitals,
$    \psi_n(k)=(
    u_{nk}, v_{nk})^T$, where $u_{nk}~ \text{and} ~ v_{nk}$ are the particle and hole sector of dimension $N \times 1$ \cite{PRXQuantum.3.040321}.\\
The Krein normalization for the positive band is given by,
\begin{align}
    \psi_n(k)^\dagger \Sigma \psi_n(k) =1, \quad
    \sum_{i=1}^N ({|u_{nk}^i|}^2- {|v_{nk}^{i}|}^2) =1
\end{align}
Thus, the probability density of the $i$ th orbital is given by $\rho^\Sigma=({|u_{nk}^i|}^2- {|v_{nk}^{i}|}^2)$, modified by the hole part, plotted in the Fig.~\ref{fig:prob_density}(a). This is different from conventional fermionic probability density.
The $\Sigma$ is the symplectic structure of the BdG space and can be connected with the symplectic structure of the position and momentum phase space via Cayley transformation \cite{1sz1-nwzf}.\\
The symplectic projector is defined as, 
\begin{align}
    P_n&=\ket{\psi_n}\bra{\psi_n}\Sigma &
    Q_n&=\mathds{1}_{2N}-P_n
\end{align}
The symplectic gauge invariant quantum metric \cite{1sz1-nwzf} characterizes the infinitesimal distance $ds^2$ between neighboring Bogoliubov modes $(\psi_n)$. In one dimension,
\begin{align}
    ds^2=1- |\psi_n^\dagger(k)\Sigma \psi_n(k+dk)|^2=g_n(k)dk^2+ \mathcal{O}(dk^3)\label{distance}
\end{align}
After simplifying Eq.~\eqref{distance} we obtain,
\begin{align}
    g_n(k)= \bra{\partial_{k}\psi_n}\Sigma Q_n\ket{\partial_{k}\psi_n}
\end{align}
\suppsection{Spectrum and eigenvectors of the dynamical matrix}

For a bosonic system, the dynamical matrix is not hermitian, so a priori it is not clear to what extent its eigenvectors and eigenvalues can be interpreted physically. In this section, we explain some important properties of its spectrum. As we argued above, for any momentum $k$, the dynamical matrix $\mathcal{D}_k$ obeys the pseudo-hermiticity relation,
\begin{align}
    \mathcal D_k^\dagger=
    \Sigma\mathcal{D}_k\Sigma.
    \label{eq:pseudo_hermiticity}
\end{align}
where we dropped the momentum subscript.
Pseudo-hermiticity alone does not guarantee a purely real spectrum.
However, since $\mathcal{D}_k = \Sigma M$ with the (hermitian) Hamiltonian matrix $M$, we will show that a sufficient condition for a real spectrum of $\mathcal D$ is the strict positive definiteness of $M_k$,
\begin{equation}
    M_k>0 \label{eq:positive_definite_condition}
    \Longleftrightarrow
    z^\dagger M_k z >0,
\end{equation}
for every nonzero vector $z\in\mathds C^{2N}$. Physically, this condition must be fulfilled because it ensures the energetic stability of the classical configuration against
fluctuations \cite{COLPA1978327, PhysRevB.100.075414, COLPA1986417, 10.1093/ptep/ptaa151}.

Let us therefore assume that $M_k$ is hermitian and positive definite. Hence, it possesses a unique positive-definite Hermitian square
root,
\begin{equation}
    \mathcal S
    =
    M_k^{1/2}
    \qquad \text{with} 
    \qquad
    \mathcal S^\dagger=\mathcal S,\quad
    \mathcal S>0,
    \label{eq:positive_square_root}
\end{equation}
Equivalently, $\mathcal{S}$ is nonsingular and satisfies $M_k=\mathcal S^{2}$. We now introduce the matrix
\begin{equation}
    \widetilde{\mathcal D}
    =
    \mathcal S \Sigma\mathcal S.
    \label{eq:hermitian_partner}
\end{equation}
This matrix is also hermitian since
\begin{align}
    \widetilde{\mathcal D}^{\dagger}
&=
    \mathcal S^\dagger
    \Sigma^{\dagger}
    \mathcal S^\dagger
=
    \mathcal S\Sigma\mathcal S
=
    \widetilde{\mathcal D}.
    \label{eq:hermitian_partner_proof}
\end{align}
Furthermore, $\widetilde{\mathcal D}$ and the dynamical matrix
$\mathcal{D}_k$ are related by the similarity transformation
\begin{align}
    \mathcal S
    \mathcal{D}_k
    \mathcal S^{-1}
&=
    \mathcal S
    \Sigma
    M_k
    \mathcal S^{-1}
=
    \mathcal S
    \Sigma
    \mathcal S^{2}
    \mathcal S^{-1}
=
    \mathcal S
    \Sigma
    \mathcal S
=
    \widetilde{\mathcal D}.
    \label{eq:similarity_transformation}
\end{align}
Therefore, $\mathcal{D}_k$ is similar to the Hermitian matrix
$\widetilde{\mathcal D}$ and thus has an identical
characteristic polynomial and hence identical eigenvalues. Since
$\widetilde{\mathcal D}$ is Hermitian, all of its eigenvalues are real, and so we find that the dynamical matrix has a real spectrum as well,
\begin{equation}
\operatorname{spec}\mathcal{D}_k\subset\mathbb R.
    \label{eq:real_spectrum_condition}
\end{equation}
Moreover, as the hermitian matrix $\widetilde{\mathcal{D}}$ is diagonalizable and $\mathcal S$ is nonsingular, the dynamical matrix
$\mathcal{D}_k$ is also diagonalizable and possesses a complete set of right eigenvectors $\ket{\psi_n}$. To demonstrate this, we start from the eigenvalue equation for $\widetilde{\mathcal D}$,
\begin{equation}
    \widetilde{\mathcal D}
    \left|w_n\right\rangle
    =
    \omega_n
    \left|w_n\right\rangle,
    \qquad
    \omega_n\in\mathbb R.
\end{equation}
Equation~\eqref{eq:similarity_transformation} now implies that the vectors $\ket{\psi_n} = \mathcal S^{-1}    \left|w_n\right\rangle$ are the corresponding right eigenvectors of $\mathcal{D}_k$,
\begin{equation}
    \mathcal{D}_k
    \ket{\psi_{n}}
=
    {\mathcal S}^{-1} \widetilde{\mathcal D} {\mathcal S} \ket{\psi_n}
=
    {\mathcal S}^{-1} \widetilde{\mathcal D} \ket{w_n}    
    =
    \omega_n
    \ket{\psi_{n}}.
\end{equation}
The orthogonality condition $\langle w_n|w_m\rangle = 0$ for $n \neq m$ implies that the eigenvectors $\ket{\psi_n}$ are mutually orthogonal with respect to the symplectic scalar product, i.e.,
\begin{align}
    \langle \psi_n|\Sigma|\psi_m\rangle = \bra{w_n} {\mathcal S}^{-1} \Sigma \mathcal{S}^{-1} \ket{w_m}
=
    \bra{w_n} \widetilde{\mathcal D}^{-1} \ket{w_m}
=
    \frac{1}{\omega_m} \langle w_n|w_m \rangle = 0
\end{align}
for $m \neq n$. Since ${\mathcal D} = \Sigma M_k$, the eigenvalue problem for the Hamiltonian matrix $M_k$ can also be expressed as the generalized eigenvalue equation
\begin{equation}
    M_k
    \ket{\psi_n}
    =
    \Sigma {\mathcal D} \ket{\psi_n} = 
    \omega_n
    \Sigma \ket{\psi_n}.
    \label{eq:generalized_eigenvalue_problem}
\end{equation}
Multiplying this equation from the left by
$\bra{\psi_n}$ gives
\begin{equation}
    \bra{\psi_n}
    M_k
    \ket{\psi_n}
    =
    \omega_n
    \bra{\psi_n}
    \Sigma \ket{\psi_n}.
    \label{eq:energy_norm_relation}
\end{equation}
As $M$ is positive-definite, the left-hand side is strictly positive for the nonzero
vector $\ket{\psi_n}$. Hence, the sign of the symplectic norm is determined by the sign of
the eigenvalue,
\begin{equation}
    \operatorname{sgn}
    \left[
        \bra{\psi_n}
    \Sigma \ket{\psi_n}
    \right]
    =
    \operatorname{sgn}(\omega_n).
    \label{eq:norm_frequency_sign}
\end{equation}
The eigenvectors of the dynamical matrix can therefore be normalized according to
\begin{equation}
    \bra{\psi_n}
    \Sigma \ket{\psi_m}
    =
    \text{sgn}(\omega_n) \delta_{nm}.
    \label{eq:symplectic_normalization}
\end{equation}
In particular, positive-frequency magnon modes have positive symplectic norm, whereas their negative-frequency particle-hole partners have negative symplectic norm. Moreover, the eigenstates $\ket{\psi_n}$ come in pairs with opposite eigenvalues and allow a completeness relation. Denoting by $\ket{\psi_n}$ the right eigenvalue of $\mathcal D$ with positive eigenvalue $\omega_n$, we denote by $\ket{\psi_n^-}$ the eigenvector with the negative eigenvalue $-\omega_n$. This is true as the dynamical matrix possesses a particle-hole constraint \cite{1sz1-nwzf}. Then, we find
\begin{align}\label{eq:app_completeness}
    \mathds{1}_{2N}= \sum_{m=1}^N \ket{\psi_m(k)}\bra{\psi_m(k)}\Sigma-\sum_{m=1}^N \ket{\psi^-_{m}(k)}\bra{\psi^-_{m}(k)}\Sigma.
\end{align}

We would like to emphasize that Eq.~\eqref{eq:positive_definite_condition} is a
sufficient but not necessary condition for a real bosonic
Bogoliubov spectrum. A more general quadratic Hamiltonian may have real eigenvalues over a finite parameter range, but the corresponding configuration is then not energetically stable. Strict positivity of $M$
provides a stronger condition, ensuring that the spectrum of the
dynamical matrix is real and that it is diagonalizable by a pseudo-unitary Bogoliubov
transformation \cite{COLPA1978327, 10.1093/ptep/ptaa151}.

\suppsection{Pinning potential}
\begin{figure}[t]
    \centering
    \includegraphics[width=0.5\linewidth]{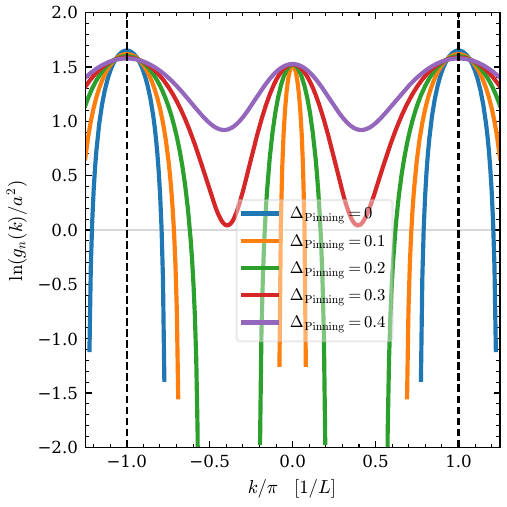}
    \caption{Quantum metric (logarithmic scale) for different values of the pinning potential applied at the edge of the unit cell. For the given physical parameters, the pinning potential must exceed a certain threshold to gap out the Goldstone mode. In the absence of a gap, the translational mode leads to a divergence of the quantum metric around $k=0$. In contrast, near the Brillouin-zone boundaries, the quantum-metric profile remains qualitatively similar for different values of the pinning potential. To regularize the calculation, we therefore choose a pinning potential sufficiently large to produce a fully gapped dispersion across the Brillouin zone.}
\end{figure}
In the absence of pinning, the continuous translational degeneracy of
the soliton lattice gives rise to a zero-frequency Goldstone mode \cite{PhysRevB.111.024411, PhysRevB.79.220405, PhysRevB.77.132405, PhysRevB.78.064425}. In
this case, the quadratic matrix is generally only positive
semidefinite, $M_k\geq 0$ at $k=0$,
and the inverse square root used in
Eq.~\eqref{eq:similarity_transformation} does not exist on the null
subspace. The above similarity transformation proof therefore applies
directly only to the nonzero-mode sector and does not, by itself,
establish diagonalizability at zero frequency ($M^{-1/2}_k$ does not exist for $k=0$). Bosonic zero modes can
require a separate canonical treatment and may be associated with
generalized eigenvectors or Jordan chains \cite{COLPA1986417}. In the
present calculation, a weak pinning potential removes the
translational zero mode. Provided that the resulting quadratic matrix
is strictly positive definite, the similarity transformation derived
above applies to the complete spectrum.
We have introduced a commensurate pinning potential \cite{PhysRevB.93.054403, PhysRevB.100.024411, doi:10.7566/JPSJ.87.033708, PhysRevB.92.220412} aligned locally with the equilibrium spin direction for $m$ magnetic unit cells to regularize the translational mode. This corresponds to the Hamiltonian
\begin{align}
   \mathcal{H}_{\text{Pinning}}= -\Delta_{\text{Pinning}} \sum_m (n_1^0\cdot \vec{S}_{m,1}+n_N^0\cdot \vec{S}_{m,N}).
\end{align}
After rotation to the $z$ axis in the local frame of reference, the term transforms to,
\begin{align}
    \mathcal{H}_{\text{Pinning}}=
    -\Delta_{\text{Pinning}} \sum_m (S'^{z}_{m,1} + S'^{z}_{m,N}).
\end{align}
Here $n_i^0= (\cos\phi_i, \sin\phi_i, 0)^T$ is the unit vector along the classical equilibrium spin direction at site $i$ and $\Delta_{\text{Pinning}} \ll J$
This gaps out the zero mode, and $M_k$ at $k=0$ can be diagonalized using the pseudo-unitary transformation.

\suppsection{wave-packet dynamics}
We construct a magnon wave packet in the $n$th nondegenerate positive-energy magnon band as a superposition of the eigenstates $\ket{\psi_n(k)}$ of the dynamical matrix,
\begin{align}
  \ket{W(t)} &= \int \frac{dk}{2\pi}\,\tw(k,t)\,e^{ikx}\,\ket{\unk},
  \label{eq:wave-packet}
  \end{align}
  
where $\tw(k,t) = |\tw(k,t)|\,e^{i\gamma(k,t)}$ is a smooth complex envelope
sharply peaked around the mean crystal momentum $\kc(t)$. For simplicity, we assumed an infinite system length such that the sum over momenta can be replaced by an integral.
We define the normalization by
\begin{align}
  \bbraket{W| W} 
&= \int dx \langle W(t)|\Sigma | W(t) \rangle 
  = \iint \frac{dk_1}{2\pi}\frac{dk_2}{2\pi}\,
    \tw^*(k_1)\tw(k_2)\,\tzinner{\psi_{n}(k_1)}{\psi_{n}(k_2)}
    \underbrace{\int dx\,e^{i(k_2-k_1)x}}_{2\pi\delta(k_2-k_1)} \notag \\
&= \int \frac{dk}{2\pi}|\tw(k)|^2 \stackrel{!}{=} 1.
  \label{eq:norm}
\end{align}
The $\Sigma$ weighted inner product ensures that we can normalize the wave packet according to Eq.~\eqref{eq:symplectic_normalization}.\\
The first moment $\langle\hat{x}\rangle$ of the position operator, which we identify with the mean position, i.e., the center of the wave packet, is given by
\begin{align}
  \xc(t) = \bbraket{W|\hat{x}|W}
  &= \iint\frac{dk_1}{2\pi}\frac{dk_2}{2\pi}\,
     \tw^*(k_1)\tw(k_2)\,\tzinner{\psi_{n}(k_1)}{\psi_{n}(k_2)}
     \int dx\; x\,e^{i(k_2-k_1)x},
     \label{eq:x1_setup}
\end{align}
Using 
\begin{equation}
  \int dx\; x\,e^{i(k_2-k_1)x} = -2\pi i \partial_{k_2}\delta(k_2-k_1).
  \label{eq:x_dist}
\end{equation}
we obtain
\begin{align}
    \xc(t)
&=
    -2 \pi i \int \frac{dk_1}{2\pi} \frac{dk_2}{2\pi} \tw^*(k_1)\tw(k_2)\bra{\psi_{n}(k_1)}\Sigma\ket{\psi_{n}(k_2)}\partial_{k_2}\delta(k_2-k_1),\label{k1} \notag \\
&=
    2 \pi i\int \frac{dk_1}{2\pi} \frac{dk_2}{2\pi} \tw^*(k_1)
    \partial_{k_2}\!\Bigl[\tw(k_2)\,\tzinner{\psi_{n}(k_1)}{\psi_{n}(k_2)}\Bigr]\,
    \delta(k_2-k_1) \notag \\
&=
    i\int \frac{dk}{2\pi}\,\tw^*(k)
    \Bigl[(\partial_k\tw(k))\underbrace{\tzinner{\unk}{\unk}}_{=1}
    + \tw(k)\,\tzinner{\unk}{\dkunk}\Bigr].
\end{align}

Writing $\tw = |\tw|e^{i\gamma}$ we have $\tw^*\partial_k\tw = |\tw|\partial_k|\tw| + i|\tw|^2\partial_k\gamma$. The first term $\propto \partial_k \abs{\tilde{w}}^2/2$ leads to a total derivative term that integrates to zero. The symplectic Berry connection is defined as,
\begin{equation}
  \mathcal{A}_{nm}(k)= i\,\tzinner{\unk}{\partial_k\psi_m(k)}, \quad \mathcal{A}_{nn}(k) \equiv \mathcal{A}_n(k) \in \mathbb{R}
  \label{eq:berry_conn}
\end{equation}
and it is real because $\partial_k\tzinner{\unk}{\psi_n(k)}=0$. Finally, we obtain the first moment
\begin{equation}
  \xc(t) = \int \frac{dk}{2\pi}\,|\tw(k,t)|^2
    \Bigl[-\partial_k\gamma(k,t) + \An(k)\Bigr].
  \label{eq:first_moment}
\end{equation}
Next, we calculate the second moment of the position operator,
\begin{align}
  \bbraket{W| \hat{x}^2| W}
  &= \iint\frac{dk_1}{2\pi}\frac{dk_2}{2\pi}\,
     \tw^*(k_1)\tw(k_2)\tzinner{\psi_n(k_1)}{\psi_n(k_2)}
     \int dx \,x^2 e^{i(k_2-k_1)x}.\label{2moment}
\end{align}
Now, the $x$ integral gives rise to the term,
\begin{equation}
  \int dx\;x^2\,e^{i(k_2-k_1)x} = -2\pi\,\partial^2_{k_2}\delta(k_2-k_1).
  \label{eq:x2_dist}
\end{equation}
After performing two integrations by parts over $k_2$ and collapsing $k_2\to k$ due to the Dirac delta function, Eq.~(\ref{2moment}) can be written as
\begin{equation}
  \bbraket{W|\hat{x}^2| W}
  = -\int \frac{dk}{2\pi}\,\tw^*\Bigl[
      \underbrace{\partial^2_k\tw}_{T_1}
      + 2\underbrace{(\partial_k\tw)\tzinner{\unk}{\dkunk}}_{T_2}
      + \underbrace{\tw\,\tzinner{\unk}{\partial^2_k\unk}}_{T_3}
    \Bigr].
  \label{eq:x2_terms}
\end{equation}
Performing an integration by parts, the first integral containing $T_1$ can be written as
\begin{equation}
  T_1 = -\int \frac{dk}{2\pi}\,\tw^*\partial^2_k\tw
  \xrightarrow{\mathrm{IBP}}
  \int \frac{dk}{2\pi}|\partial_k\tw|^2
  = \int \frac{dk}{2\pi}\!\left[(\partial_k|\tw|)^2 + |\tw|^2(\partial_k\gamma)^2\right].
\end{equation}
Using $\tzinner{\unk}{\dkunk}=-i\An$ and $\tw^*\partial_k\tw = |\tw|\partial_k|\tw|+i|\tw|^2\partial_k\gamma$,
then integrating $|\tw|\partial_k|\tw|=\tfrac{1}{2}\partial_k|\tw|^2$
by parts, the second integral containing $T_2$ can be written as
\begin{equation}
  T_2 = \int \frac{dk}{2\pi}\,|\tw|^2\!\left[-i\partial_k\An - 2\An\partial_k\gamma\right].
\end{equation}
From the completeness relation \eqref{eq:app_completeness}, we can define the projectors for positive and negative subspaces as:
\begin{align}
  P_+ &= \sum_{m=1}^N \ket{\psi_m(k)}\bra{\psi_m(k)}\Sigma \notag \\
  P_- &= -\sum_{m=1}^N \ket{\psi_m^-(k)}\bra{\psi_m^-(k)}\Sigma,
\end{align}
Then, we have $\bra{\psi_n}\Sigma P_+ = \bra{\psi_n} \Sigma$ and $\bra{\psi_n}\Sigma P_-=0$. By projecting the gradient of the eigenstate on the positive-energy subspace, we obtain
\begin{align}
    P_+\ket{\partial_k \psi_n(k)}
&= 
    \ket{\psi_n(k)}\bra{\psi_n(k)}\Sigma\ket{\partial_k \psi_n(k)}
     + 
    \sum_{m\neq n}\ket{\psi_m(k)}\bra{\psi_m(k)}\Sigma\ket{\partial_k \psi_n(k)}. 
\end{align}
and therefore,
\begin{align}
    \bra{\partial_k\unk}\Sigma P_+\ket{\partial_k \psi_n(k)} 
&= 
    \mathcal{A}_n^2+\sum_{m \neq n} \abs{\mathcal{A}_{nm}}^2.
    \label{psub}
\end{align}
Analogously, we find for the negative-energy subspace,
\begin{align}
     \bra{\partial_k\unk}\Sigma P_-\ket{\partial_k \psi_n(k)}&= -\sum_{m=1}^N \bra{\partial_k\unk}\Sigma \ket{\psi_m^-(k)}\bra{\psi_m^-(k)}\Sigma \ket{\partial_k\unk}= -\sum_{m=1}^N \abs{\mathcal{C}_{nm}}^2 \label{nsub}.
 \end{align}
where the matrix $\mathcal{C}_{nm}= \bra{\psi_m^-(k)}\Sigma \ket{\partial_k \psi_n(k)}$ quantifies the mixing between positive and negative subspaces. Now, incorporating the completeness relation and using Eq.~(\ref{psub}) and Eq.~(\ref{nsub}), we obtain
\begin{align}
    \bra{\partial_k\unk}\Sigma\ket{\partial_k \psi_n(k)} 
&= 
    \bra{\partial_k\unk}\Sigma(P_++ P_-)\ket{\partial_k \psi_n(k)}
=
    \mathcal{A}_n^2+\sum_{m \neq n} \abs{\mathcal{A}_{nm}}^2 -\sum_{m=1}^N \abs{\mathcal{C}_{nm}}^2.
\end{align}
Differentiating $\bra{\psi_n(k)} \Sigma \ket{\partial_k \psi_n(k)}=-i \mathcal{A}_n$ with respect to $k$, we obtain
\begin{align}
    \bra{\unk}\Sigma\ket{\partial_k^2 \unk} 
&= 
    -i \partial_k \mathcal{A}_n- \bra{\partial_k\unk}\Sigma\ket{\partial_k \psi_n(k)},\label{T3ref} \notag \\
&= 
    -i \partial_k \mathcal{A}_n -\mathcal{A}_n^2-\sum_{m \neq n} \abs{\mathcal{A}_{nm}}^2+\sum_{m=1}^N \abs{\mathcal{C}_{nm}}^2.
\end{align}
Combining these results, the $T_3$ integral can be expressed as:
\begin{align}
    T_3 
&= 
    \int \frac{dk}{2\pi}\,|\tw|^2
    \Bigl[i\partial_k \An + \An^2+\sum_{m\neq n} \abs{\mathcal{A}_{nm}}^2-\sum_{m=1}^N \abs{\mathcal{C}_{nm}}^2\Bigr] \notag \\
&=
    \int \frac{dk}{2\pi}\,|\tw|^2
    \Bigl[i\partial_k \An + \bra{\partial_k \psi_n(k)}\Sigma\ket{\partial_k \psi_n(k)}\Bigr].
\end{align}
Adding the three contributions $T_1$, $T_2$ and $T_3$, we obtain,
\begin{align}
    \bbraket{W|\hat{x}^2 |W}
&=
    \sigma^2_{\text{env}} +\int \frac{dk}{2\pi}\,|\tw|^2
    \!\left[(\An-\partial_k\gamma)^2 + \gnn(k)\right],
  \label{eq:second_moment} \notag \\
    \sigma_{\text{env}}^2
&=
    \int \frac{dk}{2\pi}\!(\partial_k|\tw|)^2
\end{align}
where the symplectic quantum metric of the $n$th magnon band is defined as
\begin{equation}
  \gnn(k)= \tzinner{\dkunk}{\dkunk} - \An^2(k) = \sum_{m \neq n} \abs{\mathcal{A}_{nm}}^2-\sum_{m=1}^N \abs{\mathcal{C}_{nm}}^2.
  \label{eq:quantum_metric}
\end{equation}
It follows that the quantum metric remains positive under weak particle-hole mixing. This is consistent with Fig.~\ref{fig:prob_density}(a), where the particle amplitudes $u_i$ dominate over the hole amplitudes $v_i$. The spread of the wave packet in real space can be written as
\begin{align}
  \Delta x^2
  &= \bbraket{W|\hat{x}^2 |W} - \xc^2
   = \sigma_{\text{env}}^2+\int \frac{dk}{2\pi}|\tw|^2(\An-\partial_k\gamma)^2
     - \xc^2
   + \int \frac{dk}{2\pi}\,|\tw|^2\,\gnn(k). \label{variance}
\end{align}
In the rigid-envelope approximation, the envelope $\tilde{w}$ does not decay over time.

For a narrow wave packet in momentum space, we can use $|\tw|^2\to 2\pi\delta(k-\kc)$. The center of the wave-packet and its spread can then be expressed as
\begin{align}
  \xc &= -\partial_k\gamma\big|_{\kc} + \An(\kc), \label{eq:xc_narrow} \\
  \Delta x^2 &\approx \sigma_{\text{env}}^2+ \gnn(\kc). \label{eq:bound}
\end{align}
Hence, $\sigma_{\text{env}}$ represents the initial spread of the wave packet in real space, as defined by the construction. Note that $\sigma_{\text{env}}$ becomes large for a sharply localized wave packet in momentum space, reflecting the uncertainty principle. The metric term adds a new contribution on top of the generic spread.

\suppsection{Rigid wave packet approximation}

The dynamical matrix acts on the Bloch states as
$\mathcal{D}_k\ket{\psi_n(k)} = \omega_n(k)\ket{\psi_n(k)}$, so that
\begin{align}
  \mathcal{D}_k\ket{W(t)} &= \int\!\frac{dk}{2\pi}\,\tw(k,t)\,
    \omega_n(k) e^{ikx}\ket{\psi_n(k)}, \label{eq:dy}\\
  i\,\partial_t\ket{W(t)} &= \int\!\frac{dk}{2\pi}\,
    i\partial_t\tw(k,t)\,e^{ikx}\ket{\psi_n(k)} . \label{eq:time}
\end{align}
From Eqs.~\eqref{eq:dy} and \eqref{eq:time}, using $\Sigma$ weighted inner product of the wave packet of the $n$th positive magnon band gives,
\begin{equation}
  i\,\partial_t\tw(k,t) = \omega_n(k)\,\tw(k,t),
  \qquad
  \tw(k,t) = \tw(k,0)\,e^{-i\omega_n(k)t} .
  \label{eq:env-evolution}
\end{equation}
Writing $\tw = |\tw|e^{i\gamma}$, Eq.~\eqref{eq:env-evolution}
separates into
\begin{equation}
  |\tw(k,t)| = |\tw(k,0)|,
  \qquad
  \gamma(k,t) = \gamma(k,0) - \omega_n(k)\,t .
  \label{eq:polar}
\end{equation}
The modulus is conserved at each $k$ independently, so all deformations of
the wave packet are carried by the phase $\gamma(k,t)$. In
particular, since
$\sigma_{\text{env}}^{2} =(2\pi)^{-1} \int\!dk(\partial_k|\tw|)^{2}$
is a functional of $|\tw|$ alone, we find
\begin{equation}
  \partial_t \sigma_{\text{env}} = 0
\end{equation}
Let us define $d\xi(k) = \frac{dk}{2\pi}|\tw(k)|^{2}$. From
Eq.~\eqref{variance} we then have,
\begin{equation}
  \langle(\Delta x)^{2}\rangle = \sigma_{\text{env}}^{2}
  + \operatorname{Var}_\xi[\,\mathcal{A}_n - \partial_k\gamma\,]
  + g_n(\kc).
\end{equation}
We introduce the $k$-resolved position
$X(k,t) \equiv \mathcal{A}_n(k) - \partial_k\gamma(k,t)$, which is
gauge invariant as $\mathcal{A}_n$ and
$\partial_k\gamma$ shift identically under
$\ket{\psi_n(k)}\to e^{i\varphi(k)}\ket{\psi_n(k)}$. Now, differentiating Eq.~\eqref{eq:polar} with respect to $k$ gives the
exact relation
\begin{equation}
  X(k,t) = X(k,0) + v_n(k)\,t,
  \qquad v_n(k) = \partial_k\omega_n(k),
  \label{eq:X-ballistic}
\end{equation}
Taking the expectation value with $d\xi(k)$, we have $x_c(t) = x_c(0) + \langle v_n\rangle t$. Now, we calculate the variance with
respect to the measure $\xi$,
\begin{equation}
  \operatorname{Var}[X(t)] = \operatorname{Var}[X_0]
  + 2t\operatorname{Cov}[X_0,v_n]
  + t^{2}\operatorname{Var}[v_n].
  \label{eq:var-split}
\end{equation}
Here we use,
\begin{align}
    \operatorname{Var}[X]&= \operatorname{E}[X^2]-(\operatorname{E}[X])^2\\
    \operatorname{Var}[X+Y]&= \operatorname{Var}[X]+2\operatorname{Cov}[X,Y]+\operatorname{Var}[Y]\\
    \operatorname{Var}[aX+b]&= a^2\operatorname{Var}[X]\\
    \operatorname{Cov}[X,Y]&=\operatorname{E}[XY]-\operatorname{E}[X]\operatorname{E}[Y], \quad \operatorname{E}[.]\equiv\int d\xi (.)
\end{align}
If $\operatorname{Var}[v_n]=0$ then from Cauchy--Schwarz inequality,
\begin{align}
  \big|\operatorname{Cov}_\xi[X_0,v_n]\big|
  &\leq \operatorname{Var}_\xi^{1/2}[X_0]\,
     \operatorname{Var}_\xi^{1/2}[v_n]=0,
\end{align}
The spread is thus time independent if and only if
$\operatorname{Var}[v_n] = 0$. The time scale for a dispersive band can be given by, $t \ll \sigma_{\text{env}}/\operatorname{Var}^{1/2}[v_n]$. If $\partial_k^2\omega_n(k) \approx 0$, we have a larger time scale for which our rigid wave packet approximation is valid.

\suppsection{External Perturbation}
We consider an effective position-dependent chemical potential by changing the local magnon density of different bands. The perturbation Hamiltonian arising from this chemical potential can be written as
\begin{equation}
    \delta \mathcal{H}=\frac{1}{2} \sum_k \Phi_k^\dagger \mu(\hat{x}) \Phi_k
    = \frac{1}{2} \sum_{k,n} \mu_n(\hat{x}) (b_{k,n}^\dagger b_{k,n}+b_{-k,n}^\dagger b_{-k,n} )
\end{equation}
Here, $\mu(x)$ is a $2N \times 2N$ dimensional diagonal matrix. The diagonal terms $\mu_n(\hat{x})$ are the chemical potentials corresponding to different bands. Now, the perturbation term for the $n$th band can be written as,
\begin{equation}
    \delta \mathcal{V}_n= \frac{1}{2}\sum_k \mu_n(\hat{x}) (b_{k,n}^\dagger b_{k,n}+b_{-k,n}^\dagger b_{-k,n} )
\end{equation}
Where, $\hat{x}=i\partial_k+\mathcal{A}_n(k)$.
Since $\mu_n(\hat{x})$ is evaluated at the operator $\hat{x}$, we Taylor expand it in
$(\hat{x} - \xc)$ around the wave packet center $\xc(t)$:
\begin{align}
  \mu_n(\hat{x})
  &= \mu_n(\xc)
   + (\hat{x}-\xc)\,\partial_{x_c} \mu_n(x_c)
   + \frac{(\hat{x}-\xc)^2}{2}\,\partial^2_{x_c} \mu_n(x_c)
   + \mathcal{O}\!\left[(\hat{x}-\xc)^3\right].
  \label{eq:B_expand}
\end{align}
The total perturbation splits as
\begin{align}
   \mu_n(\hat{x}) &= \mu_n^{(0)}
                    + \mu_n^{(1)}(\hat{x})
                    + \mu^{(2)}_n(\hat{x})
                    + \cdots, \label{eq:dH_split}
\end{align}
where
\begin{align}
   \mu_n^{(0)}
  &= \mu_n(\xc),
  \label{eq:dH0}\\[4pt]
   \mu_n^{(1)}(\hat{x})
  &= \left[(\hat{x}-\xc)\,\partial_{x_c} \mu_n(\xc)\right],
  \label{eq:dH1}\\[4pt]
   \mu_n^{(2)}(\hat{x})
  &= \left[\frac{1}{2}\,(\hat{x}-\xc)^2\,\partial^2_{x_c} \mu_n(\xc)\right].
  \label{eq:dH2}
\end{align}
\subsection{Expectation values in the wave packet}

We compute $\bbraket{W|\mu_n(\hat{x})| W}$ up to second order using the
symplectic inner product of the wave packet. The zeroth-order contribution for the positive energy band is given by
\begin{equation}
  \langle \mu_n^{(0)}\rangle
  = \mu_n(\xc)\,\underbrace{\bbraket{W|W}}_{=+1}
  = \mu_n(\xc).
  \label{eq:EV0}
\end{equation}
This term shifts the energy of each band by $\mu_n$. The first-order term can be written as
\begin{align}
  \langle \mu_n^{(1)}\rangle
  &= \partial_{x_c} \mu_n(\xc)\,
     \underbrace{\bbraket{W|(\hat{x}-\xc)|W}}_{\displaystyle=\,\langle\hat{x}\rangle - \xc\,=\,0}.
  \label{eq:EV1}
\end{align}
By definition of $\xc = \langle\hat{x}\rangle$, the first-order term vanishes identically. Using Eq.~(\ref{eq:bound}), we obtain the second-order term as
\begin{align}
  \langle \mu_n^{(2)}\rangle
  &= \frac{1}{2}\,\partial^2_{x_c} \mu_n(\xc)\,
     \bbraket{W|(\hat{x}-\xc)^2 | W},
  \label{eq:EV2_start}
\end{align}
Now using
\begin{align}
  \bbraket{W|(\hat{x}-\xc)^2| W}
  &= \langle(\hat{x}-\xc)^2\rangle
  = \sigma_{\text{env}}^2 + \gnn(\kc),
  \label{eq:x2_ev}
\end{align}
we find
\begin{align}
  \langle \mu_n^{(2)}\rangle
  &= \frac{1}{2}\,(\gnn(\kc)+\sigma_{\text{env}}^2)\,\partial^2_{x_c} \mu_n(\xc).
  \label{eq:EV2}
\end{align}
This is the central result of the perturbation theory: the quantum
metric $\gnn(\kc)$ directly controls how strongly the magnon wave packet
responds to curvature $\partial_x^2 \mu_n(x)$ of the effective potential.

\suppsection{Time-Dependent wave-packet Lagrangian}

\subsection{Dirac-Frenkel variational principle}

The semiclassical dynamics of the magnon wave packet are governed by the Dirac-Frenkel variational principle with the action
\begin{equation}
  S[\xc,\kc] = \int dt\;\mathcal{L}(\xc,\kc,\dot{x}_c,\dot{k}_c,t),
  \label{eq:action}
\end{equation}
where the Lagrangian is 
\begin{equation}
  \mathcal{L} = \bbraket{W|i\partial_t| W}
               - \bbraket{W|\mathcal{D}| W}
               - \bbraket{W|\mu_n(\hat{x})| W}.
  \label{eq:L_def}
\end{equation}
The first term is the Berry phase (kinetic) term, which generates the time derivatives of the wave-packet center in position and momentum. Evaluating $\bbraket{W|i\partial_t W}$ in the narrow-envelope limit
using the parametrization $\tw(k,t) = |\tw|e^{i\gamma(k,t)}$ and
$\xc = -\partial_k\gamma|_{\kc} + \An(\kc)$, one finds
\begin{align}
  \bbraket{W|i\partial_t W}
  &=-\frac{d \gamma}{dt}+\dot{k}_c(-\xc + \An(\kc))
  = \kc\dot{x}_c + \dot{k}_c\An(\kc)-\frac{d}{dt}(x_c k_c)-\frac{d\gamma}{dt} \doteq \mathcal{L}_t\\
  \mathcal{L}_t &\doteq \kc\dot{x}_c + \dot{k}_c\An(\kc),
  \label{eq:kinematic}
\end{align}
where $\doteq$ means equality up to a total time derivative.
The second term in the Lagrangian is the expectation of the unperturbed Hamiltonian with respect to the wave packet. In the narrow-envelope limit, we have
\begin{equation}
  \bbraket{W|\mathcal{D}| W} = \En(\kc).
  \label{eq:E0}
\end{equation}
The third term is from the external perturbation. It acts as an operator on the wave packet. Using the previous results, we obtain
\begin{align}
  \bbraket{W|\mu_n(\hat{x})| W}
  &= \mu_n(\xc)
   + \frac{1}{2}\,[\gnn(\kc)+\sigma_{\text{env}}^2]\,\partial^2_{x_c} \mu_n(\xc)
   + \mathcal{O}((x-\xc)^3).
  \label{eq:E_pert}
\end{align}
In summary, we obtain the full time-dependent Lagrangian using \cref{eq:kinematic,eq:E0,eq:E_pert}
\begin{align}
  \mathcal{L}(\xc,\kc,\dot{x}_c,\dot{k}_c)
  &= \kc\dot{x}_c
   + \dot{k}_c\,\An(\kc)
   - \En(\kc)
   -  \mu_n(\xc)
   - \frac{1}{2}\,(\gnn(\kc)+\sigma_{\text{env}}^2)\,\partial^2_{x_c} \mu_n(\xc).
  \label{eq:Lagrangian}
\end{align}

\suppsection{Equations of Motion for the Magnon wave-packet center}

\subsection{Euler-Lagrange equations}

Requiring the action $S=\int dt\,\mathcal{L}$ to be stationary with respect to the independent variations of $\xc(t)$ and $\kc(t)$ yields the Euler-Lagrange (EL) equations,
\begin{equation}
  \frac{d}{dt}\frac{\partial\mathcal{L}}{\partial\dot{x}_c}
  - \frac{\partial\mathcal{L}}{\partial\xc} = 0,
  \qquad
  \frac{d}{dt}\frac{\partial\mathcal{L}}{\partial\dot{k}_c}
  - \frac{\partial\mathcal{L}}{\partial\kc} = 0.
  \label{eq:EL}
\end{equation}
Using the EL equation for $x_c$, from the Lagrangian Eq.~\eqref{eq:Lagrangian}, we find
\begin{align}
  \frac{\partial\mathcal{L}}{\partial\dot{x}_c} &= \kc,
  \qquad
  \frac{d}{dt}(\kc) = \dot{k}_c.
  \label{eq:dL_dxcdot}
\end{align}
The $\xc$-derivative is given by
\begin{equation}
  \frac{\partial\mathcal{L}}{\partial\xc}
  = -\partial_{x_c} \mu_n(\xc)
    - \frac{1}{2}\,(\gnn(\kc)+ \sigma_{\text{env}}^2)\,\partial^3_{x_c} \mu_n(\xc).
  \label{eq:dL_dxc}
\end{equation}
Setting Eq.~\eqref{eq:dL_dxcdot} equal to Eq.~\eqref{eq:dL_dxc} leads to
\begin{equation}
  \dot{k}_c
  = -\partial_{x_c} \mu_n(\xc)
    - \frac{1}{2}\,(\gnn(\kc)+ \sigma_{\text{env}}^2)\,\partial^3_{x_c} \mu_n(\xc).
  \label{eq:kc_eom}
\end{equation}
The first term is the conventional force generated by the spatial gradient of the effective chemical potential. The second term is a geometric higher-gradient
force weighted by the quantum metric $\gnn$.\\

Similarly, from the Euler-Lagrange equation for $k_c$, we obtain
\begin{align}
  \frac{\partial\mathcal{L}}{\partial\dot{k}_c} &= \An(\kc),
  \qquad
  \frac{d}{dt}\An(\kc)
  = \,\partial_{k_c}\An(\kc)\,\dot{k}_c
  \label{eq:dL_dkcdot}
\end{align}
Differentiating with respect to $\kc$ yields
\begin{align}
  \frac{\partial\mathcal{L}}{\partial\kc}
  &= \dot{x}_c
   - \partial_{k_c}\En(\kc)
   - \frac{1}{2}\,\partial_{k_c}\gnn(\kc)\,\partial^2_{x_c} \mu_n(\xc,t)+\,\partial_{k_c}\An(\kc)\,\dot{k}_c.
  \label{eq:dL_dkc}
\end{align}
Setting Eq.~\eqref{eq:dL_dkcdot} equal to Eq.~\eqref{eq:dL_dkc} and solving for $\dot{x}_c$, we find
\begin{align}
  \dot{x}_c
  &= \partial_{k_c}\En(\kc)
   + \frac{1}{2}\,\partial_{k_c}\gnn(\kc)\,\partial^2_{x_c} \mu_n(\xc).
  \label{eq:xc_eom}
\end{align}
The first term is the ordinary group velocity of the $n$th band. The second term is the quantum-metric correction arising from nontrivial band geometry. Therefore, the dynamics of the wave-packet center are given by, 
\begin{align}
  \dot{x}_c
  &= \partial_k\omega_n(\kc)
   + \frac{1}{2}\,\partial_k\gnn(\kc) \partial^2_x \mu_n(\xc),
  \label{eq:xc_final}\\
  \dot{k}_c
  &= -\partial_x\mu_n(\xc)
   -\,\frac{1}{2}\,[\gnn(\kc)+\sigma_{\text{env}}^2]\,\partial^3_x \mu_n(\xc)
  \label{eq:kc_final}
\end{align}

\suppsection{Closed-form solutions}
To obtain the trajectory of the center of the wave-packet, we expand the perturbing chemical potential around the soliton center $x_0$ inside one unit cell,
\begin{align}
    \mu_n(x_c)=\mu_n(x_0)+(x_c-x_0)\frac{\partial \mu_n}{\partial x_0}+\frac{(x_c-x_0)^2}{2}\frac{\partial^2 \mu_n}{\partial x_0^2}
    +\mathcal{O}((x_c-x_0)^3), \label{eqdelB}
\end{align}
We choose the chemical potential to be maximal at the center of the soliton, so that $\partial \mu_n/\partial x_0=0,~ \partial^2 \mu_n/ \partial x_0^2 < 0$. Then Eq.~\eqref{eqdelB} becomes
\begin{align}
    \mu_n(x_c) &\approx \mu_n(x_0)-\abs{\frac{\partial^2 \mu_n}{\partial x_0^2}}\frac{(x_c-x_0)^2}{2}\\
    \mu_n(x_c) &\approx \mu_n(x_0)-\frac{\mu_n^0(x_c-x_0)^2}{2} , \quad \mu_n^0 \frac{a^2}{J} = \frac{a^2}{J}\abs{\frac{\partial^2 \mu_n}{\partial x_0^2}} \ll 1 \label{field}
\end{align}
For the dynamics, we focus on the lowest magnon band, which is nearly flat and also easy to excite at low temperature due to the bosonic nature of the magnons \cite{PhysRevResearch.1.032023, giesen2026resonantexcitationslowfrequency}. We start with the initial value of position and momentum of the wave packet center as $(x_c, k_c) = (x_0, k_I)$. We position the initial wave-packet center at the soliton center $(x_0)$ and choose the momentum $(k_I=k_0+\delta)$, where the metric $g(k_0)$ is minimum, $\partial g(k_0)/ \partial k_0=0$.\\

Using Eq.~\eqref{field}, the dynamics of the center of the magnon wave-packet can be written as,
\begin{align}
    \frac{dk_c}{dt}= \mu_n^0(x_c-x_0), \quad
    \frac{d\xc}{dt}=\frac{d\omega}{d\kc}-\frac{\mu_n^0}{2}\frac{\partial g}{\partial \kc} \label{xckcdiff}
\end{align}
Eliminating $t$ between the two equations and integrating, we obtain
\begin{align}
    \int_{x_I}^{\xc}d\xc~ \mu_n^0(\xc-x_0) = \int_{k_I}^{\kc} d\kc\left[\frac{d\omega}{d\kc}-\frac{\mu_n^0}{2}\frac{\partial g}{\partial \kc}\right]
\end{align}
Since the lowest-energy magnon band depends weakly on momentum, $\omega(k_c)\approx\omega(k_I)$, we further find
    \begin{gather}
    \xc= x_0\pm \sqrt{g(k_I)-g(\kc)}, \quad |\xc-x_0|_{\text{max}}= \sqrt{g(k_I)-g_{\text{min}}}
\end{gather}
Hence, from Eq.~\eqref{xckcdiff}, we obtain
\begin{align}
    \frac{d^2k_c}{dt^2}+\frac{\partial}{\partial k_c}\left(\frac{(\mu_n^0)^2}{2}g(k_c)-\mu_n^0\omega(k_c)\right)=0 \label{momforce}
\end{align}
We can compare Eq.~\eqref{momforce} with the equation of motion of a point particle in real space in the presence of a potential. Here, the equivalent effective potential in momentum space can be expressed as $(\mu_n^0)^2g(k_c)/2-\mu_n^0 \omega(k_c)$. This effective potential can be either attractive or repulsive, depending on contributions from the quantum metric and group velocity. \\
In the nearly flat band limit with a dominated quantum metric contribution, Eq.~\eqref{momforce} can be written as,
\begin{equation}
    \frac{d^2k_c}{dt^2}+\frac{(\mu_n^0)^2}{2}\frac{\partial g(k_c)}{\partial k_c}=0 \label{metricforce}
\end{equation}
From Eq.~\eqref{metricforce}, we can compare with motion inside a real-space attractive potential. The effective potential in momentum space is given by $(\mu_n^0)^2g(k_c)/2$, which is intrinsic to the band geometry and not equal to the effective external potential on the wave-packet. The symplectic quantum metric generates an effective potential landscape for the magnon wave packet, which is intrinsic to the band geometry and can be controlled by the curvature of the external perturbation $\mu_n^0$. \\ 
If the motion of the wave-packet is confined to the vicinity of a local minimum $k_0$ of the quantum metric, then
\begin{align}
    g'(k_0) &=0, \quad g''(k_0)>0\\
    g(k_c) &=
    g(k_0)+\cancelto{0}{\frac{\partial g(k_0)}{\partial k_0}}(k_c-k_0)+\frac{1}{2\!}\frac{\partial ^2g(k_0)}{\partial k_0^2}(k_c-k_0)^2+\mathcal{O}\left((k_c-k_0)^3\right)
    \end{align}
Equation~\eqref{metricforce} reduces to
    \begin{align}
    \frac{d^2k'}{dt^2}+\frac{(\mu_n^0)^2}{2}g''(k_0)k' &=0,& \quad k' &=(k_c-k_0)\\
    \Omega &= \sqrt{\frac{(\mu_n^0)^2}{2}g''(k_0)},& \quad T &=\frac{2\pi\sqrt{2}}{\mu_n^0\sqrt{g''(k_0)}}
\end{align}
The force equation is given by,
\begin{align}
    \frac{dk_c}{dt} &=\pm \mu_n^0\sqrt{g(k_I)-g(k_c)}
    \intertext{Since the dynamics is bounded around the minimum of the metric, we have}
    k_I=k_0+\delta, \quad g(k_I) &=g(k_0+\delta)=
    g(k_0)+\delta\cancelto{0}{\frac{\partial g(k_0)}{\partial k_0}}+\frac{\delta^2}{2\!}\frac{\partial ^2g(k_0)}{\partial k_0^2} +\mathcal{O}\left(\delta^3\right)\\
    \frac{dk_c}{dt} &= \pm\mu_n^0\sqrt{\frac{1}{2}\frac{\partial ^2g(k_0)}{\partial k_0^2}(\delta^2-(k_c-k_0)^2)}\\
    \frac{dk_c}{dt} &=\pm\Omega\sqrt{(\delta^2-(k_c-k_0)^2)}\\
    \frac{dz}{dt} &=\pm \Omega\sqrt{\delta^2-z^2}, \quad z=k_c-k_0, \quad k_0-\delta\leq k_c \leq k_0+\delta, \quad -\delta \leq z \leq \delta\\
\intertext{If the initial mean momentum of the wave-packet is $k_0+\delta$, then}
\int_{\delta}^{z}\frac{dz}{\sqrt{\delta^2-z^2}} &=-\Omega\int_0^t dt,\quad  z=\delta\cos(\Omega t), \quad k_c(t)=k_0+\delta\cos(\Omega t)\\
\frac{dk_c}{dt} &=\mu_n^0(x_c-x_0),\quad x_c(t)=x_0-\frac{\Omega \delta}{\mu_n^0}\sin(\Omega t)
\end{align}
Then the phase space motion can be written as,
\begin{align}
    x_c(t) =x_0-\frac{\Omega \delta}{\mu_n^0}\sin(\Omega t), \quad
    k_c(t) =k_0+\delta\cos(\Omega t)
\end{align}
leading to
\begin{gather}
    \frac{(x_c(t)-x_0)^2}{c_x^2}+\frac{(k_c(t)-k_0)^2}{c_k^2}=1, \quad c_x= \frac{\Omega \delta}{\mu_n^0}, \quad c_k=\delta
\end{gather}
The enclosed phase space area is given by, $A_{\text{phase}}= \pi c_x c_k= \pi \delta^2 \sqrt{g''(k_0)/2}$, which is independent of the external perturbation strength. It depends only on the initial momentum offset of the wave-packet center and the intrinsic symplectic quantum metric.
\end{document}